\documentclass[useAMS,usenatbib]{mn2e}
\pdfoutput=1
\usepackage{graphicx}
\usepackage{subfigure}
\usepackage{ifpdf}
\usepackage{amsmath}
\usepackage{hyperref}
\newcommand\msol{{{\cal M}_\odot}}
\title[GC Assembly in the Milky Way]{The Bifurcated Age-Metallicity Relation of Milky Way Globular Clusters and its Implications For the Accretion History of the Galaxy}
\author[Leaman, VandenBerg, \& Mendel]{Ryan Leaman$^{1,2}$\thanks{E-mail:
rleaman@iac.es}, Don A. VandenBerg$^{3}$, and J. Trevor Mendel$^{4}$\\
$^{1}$Instituto de Astrof\'{i}sica de Canarias, E-38205 La Laguna, Tenerife, Spain, $^{2}$Dept. Astrof\'{i}sica, Universidad de La Laguna, E-38205 La Laguna, Tenerife, Spain\\ $^{3}$Dept. of Physics \& Astronomy, University of Victoria, P.O. Box 3055, Victoria, B.C., V8W 3P6, Canada, $^{4}$Max-Planck-Institut f\"{u}r extraterrestrische Physik, Giessenbachstrasse, D-85748 Garching, Germany}
\begin{document}
\newcounter{tabcount}
\date{Accepted 2013 August 14. Received 2013 May 17}

\pagerange{\pageref{firstpage}--\pageref{lastpage}} \pubyear{2013}

\maketitle

\label{firstpage}

\begin{abstract}
We use recently derived ages for 61 Milky Way (MW) globular clusters (GCs) 
to show that their age-metallicity relation (AMR) can be divided into 
two distinct, parallel sequences at [Fe/H] $\ga -1.8$.  
Approximately one-third of the clusters form an offset sequence that spans the
full range in age ($\sim 10.5$--13 Gyr), 
but is more metal rich at a given age by
$\sim 0.6$ dex in [Fe/H].  All but one of the clusters in the offset sequence
show orbital properties that are consistent with membership in the MW
disk.  They are not simply the most metal-rich GCs, which have long been known
to have disk-like kinematics, but they are the most metal-rich clusters at all
ages.  The slope of
the mass-metallicity relation (MMR) for galaxies implies that the
offset in metallicity of the two branches of the AMR corresponds to a mass
decrement of 2 dex, suggesting host galaxy masses of $M_{*} \sim 10^{7-8} \msol$
for GCs that belong to the more metal-poor AMR.
We suggest that the metal-rich branch of the AMR consists of clusters 
that formed in-situ in the disk, while the metal-poor GCs were formed 
in relatively low-mass (dwarf) galaxies and later accreted by the MW.
The observed AMR of MW disk stars, and of the LMC, SMC and WLM dwarf 
galaxies are shown to be consistent with this interpretation, 
and the relative distribution of implied progenitor masses 
for the halo GC clusters is in excellent agreement with the 
MW subhalo mass function predicted by simulations.  A notable implication
of the bifurcated AMR, is that the identical mean ages and spread in ages,
for the metal rich and metal poor GCs are difficult to reconcile
with an in-situ formation for the latter population.
\end{abstract}

\begin{keywords}
Galaxy: formation -- (Galaxy:) globular clusters: general -- galaxies: dwarf.
\end{keywords}

\section{Introduction}
The properties of globular cluster (GC) systems in disk and elliptical galaxies
provide an opportunity not only to study the formation of these dense star
clusters, but also to probe the dynamics, chemical evolution, and assembly
history of their host galaxies (e.g., \citealt{Pota13}).  In our own Milky Way
(MW), the GCs have long been used to support different scenarios for the mass
growth of the stellar halo.  For instance, the fact that the distribution of
cluster metallicities does not vary significantly with Galactocentric distance
(R$_{\rm G}$) in GCs beyond R$_{\rm G} \sim 8$ kpc was used in early work by
\cite{SZ78} to argue against the monolithic collapse model for the formation
of the MW envisioned by \cite{ELS62}.

\begin{figure*}
\begin{center}
\ifpdf
\includegraphics[width=0.79\textwidth]{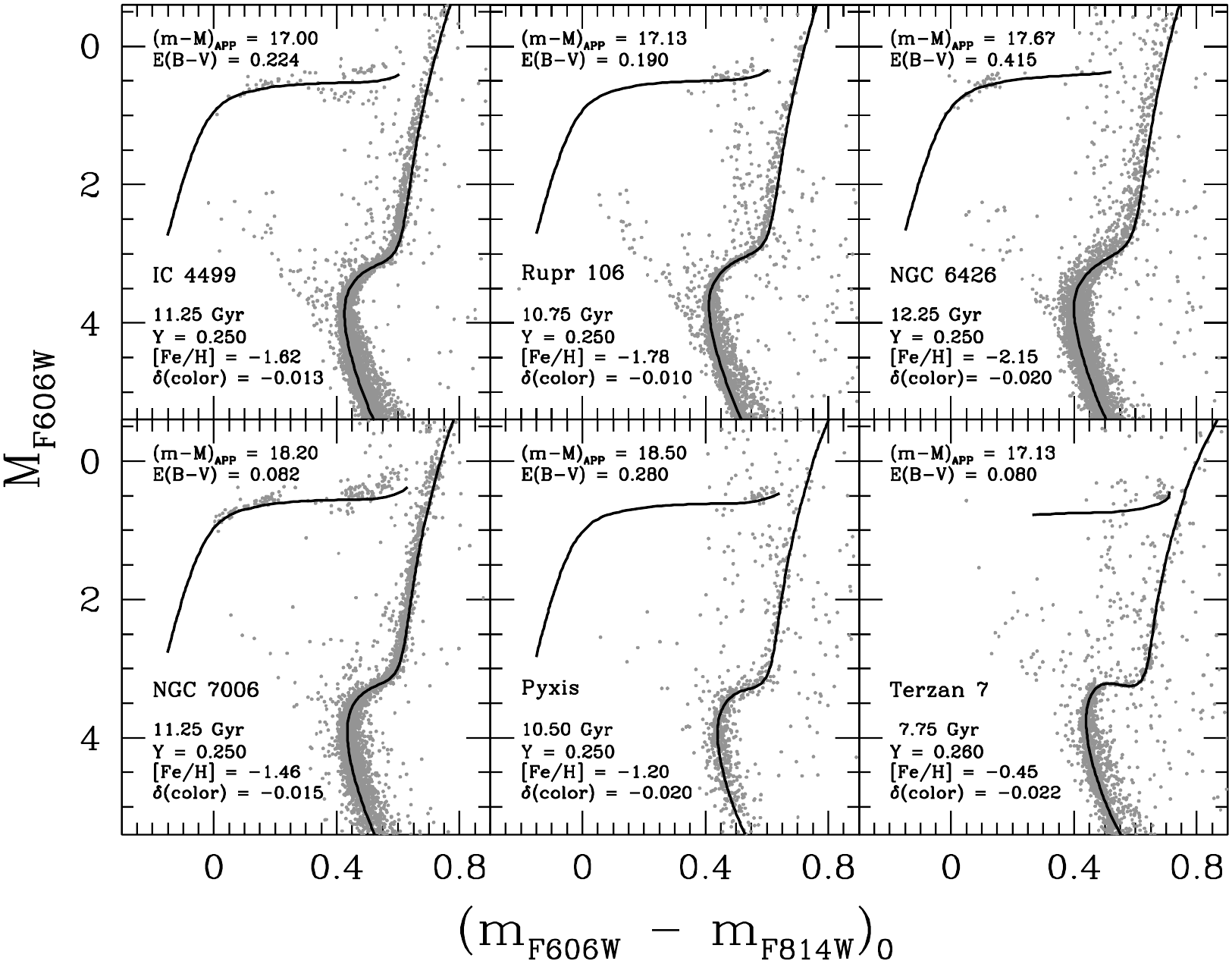}
\else
\includegraphics[width=0.79\textwidth]{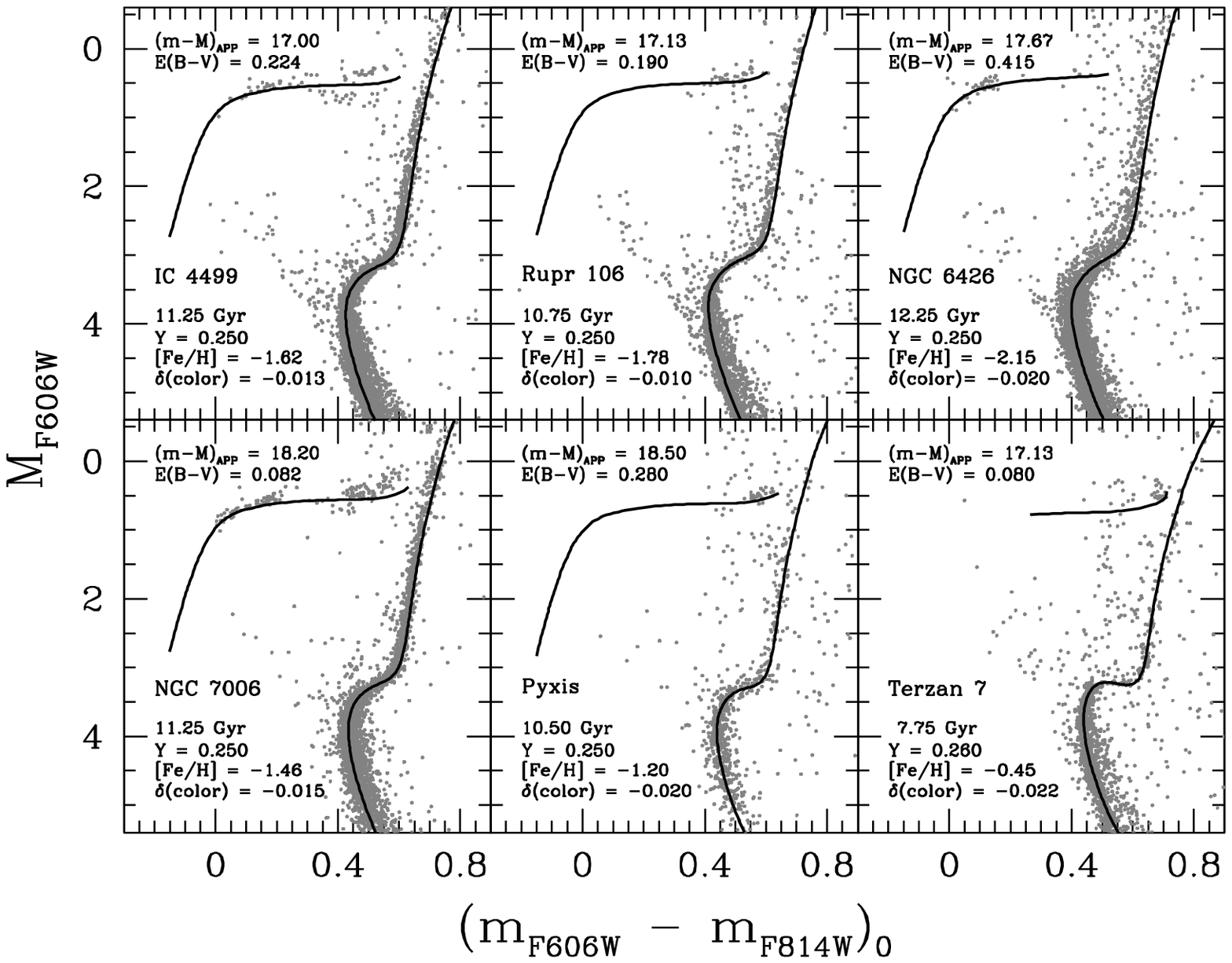}
\fi
\caption{CMDs of six outer halo globular clusters, along with the best fitting isochrones following the method described in the text and V13.  Each panel lists the adopted reddenings and apparent distance moduli that were found by fitting the ZAHB to the lower bound of the HB stars, along with the derived age.  In addition, the assumed helium and [Fe/H] values, as well as the adjustment in colour that was needed in order for the isochrone to match the observed turnoff colour are noted.}
\label{fig:ohalo}
\end{center}
\end{figure*}

As most GCs are relatively simple stellar populations that formed nearly
instantaneously, their ages are an even more useful property for investigating
such scenarios.  This is complicated, however, by the long-standing difficulty
of determining accurate relative (let alone absolute) ages for GCs (e.g.,
compare the sometimes discordant results reported by \citealt{Rosenberg99},
\citealt{VandenBerg00}, \citealt{DeAngeli05}, \citealt{MF09}).  On the one hand, current
predictions of turnoff luminosity versus age relations should be quite accurate
given the steady improvement in the basic physics ingredients of stellar models
(such as opacities and nuclear reaction rates) and the incorporation of
diffusive processes that are thought to be important in the
evolution of old, metal-deficient stars.  On the other hand, it is still very
risky to place a similar reliance on detailed fits of isochrones to the observed
colour-magnitude diagrams (CMDs) of GCs because the predicted $T_{\rm eff}$ and
colour scales have significant uncertainties associated with them (due to
inadequacies in, e.g., the treatment of super-adiabatic convection and the
atmospheric boundary condition).  Although distance information is generally
not needed when estimating relative GC ages, the latter do depend on the assumed
[Fe/H] values and any cluster-to-cluster variations that happen to be present
in the abundances of helium, oxygen, and/or other $\alpha$-elements (see
\citealt[hereafter V13, see sections 5.3 and 6.1.1]{V13}).  In fact, it has
become quite clear from recent observational studies that such variations exist
(e.g., \citealt{CG09b}), coupled with various chemical abundance
anticorrelations and bimodalities, and that they account for some of the
peculiar CMD morphologies which have been discovered (see the review by
\citealt{Gratton12}).

The latest investigation of GC ages, by V13, employed an improved version of
the classic $\Delta\,V_{TO}^{HB}$ method, which uses the magnitude difference
between the horizontal branch (HB) and the turnoff (TO) as the primary age
constraint.  The main advantage of this technique over all others that have
been proposed is that the effects of metal abundance uncertainties are minimized
--- because the luminosity of the HB and of the turnoff, at a fixed age, are
both reduced when the metallicity is increased (and vice versa), although not by
exactly the same amount.  In 
addition, V13 used what appears to be a particularly robust way of determining
the age once the absolute magnitude scale had been set by matching a theoretical
zero-age horizontal branch (ZAHB) locus for the latest estimate of the cluster
metallicity (from \citealt{CG09}) to the lower bound of the observed
distribution of its HB stars.  Because the predicted variation of $M_V$(HB) with
[Fe/H] is in very good agreement with empirical determinations (see V13), the 
age-metallicity relation (AMR) that was derived in that study should be 
especially reliable.  Interestingly, that AMR gives the visual impression of
being bifurcated at [Fe/H] $\ga -1.8$, such that clusters with halo-type or
disk-like orbits populate different branches.  In this paper, the V13 AMR has
been augmented by data for six additional outer-halo GCs (with $15 \leq$
R$_{\rm G} \leq 40$ kpc), and after carrying out an examination of the kinematic 
properties of the GCs in the two sequences, we explore some the implications of
the split AMR for the assembly of the MW.
 
\section{Globular Cluster Age Data}
We rely on the ages that V13 recently derived for the 55 Galactic GCs that they
considered (based on photometry from \citealt{Sarajedini07}), along with their adopted [Fe/H] values (from the compilation given
by \citealt{CG09}).  Since similar {\it HST} Advanced Camera for Surveys (ACS)
photometry has recently become available for a few additional outer-halo GCs
(\citealt{Dotter11})\footnote{\url{
http://www.astro.ufl.edu/~ata/public_hstgc/databases.html}}
we decided to take
this opportunity to determine their ages using exactly the same procedure and
stellar models that were employed by V13, and to include them in the AMR that
is the subject of this investigation.  Figure \ref{fig:ohalo} illustrates the 
adopted fits of ZAHB models to the cluster HB stars and of isochrones to the
turnoff (TO) portions of the observed CMDs.  Reference should be made to V13
for a thorough discussion of the fitting procedure, but the main point to be
emphasized here is that the age of each cluster is based on the fit of
isochrones only to the observations in the vicinity of the turnoff (i.e., to
the stars that have colours within $\simeq 0.05$ mag of the TO colour), where
the morphology of the isochrones is predicted to be independent of the age and
nearly independent of [Fe/H], [$\alpha$/Fe], helium abundance 
and the mixing-length parameter.

\begin{figure*}
\centering
\mbox{\subfigure{\ifpdf
\includegraphics[width=90mm]{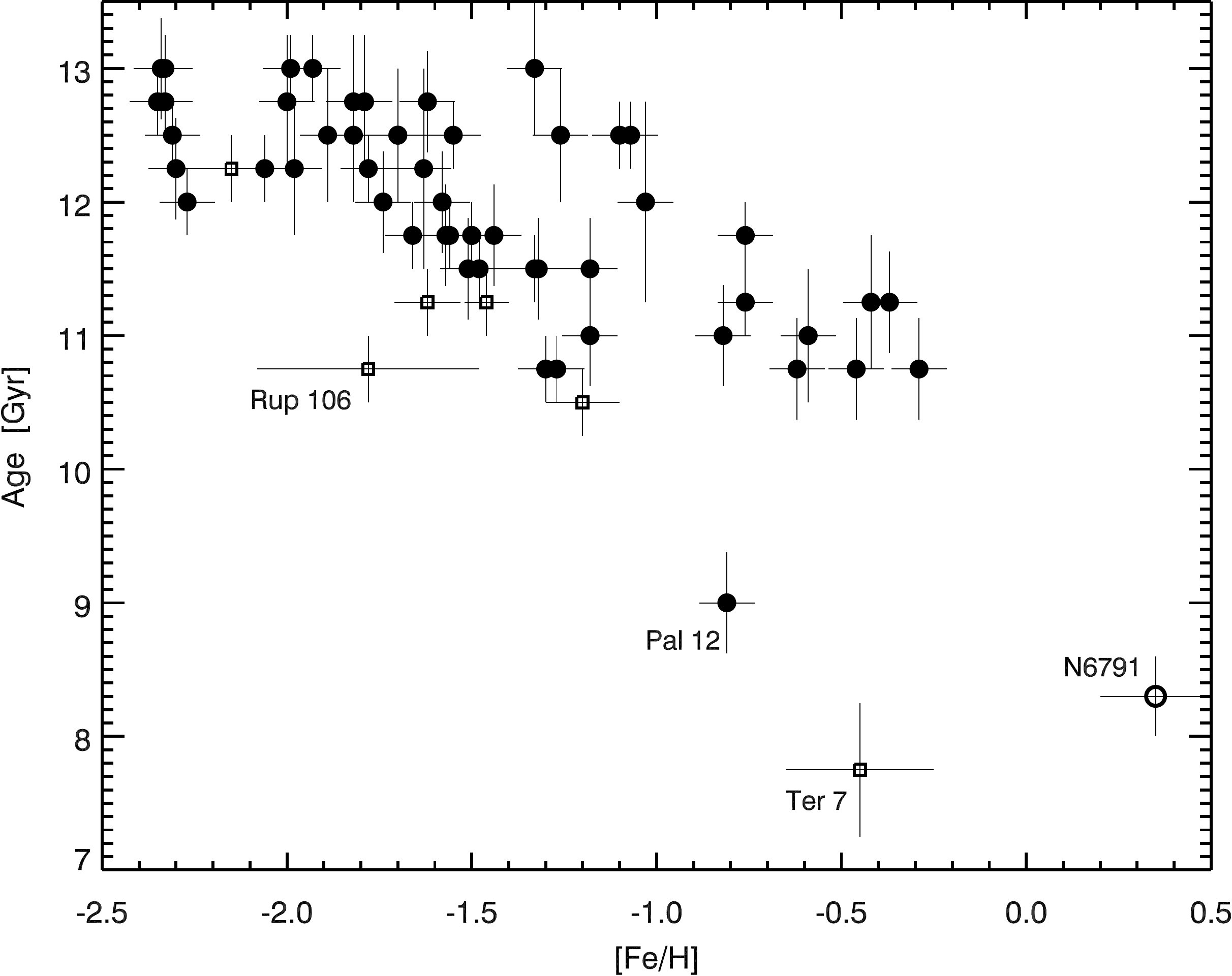}
\else
\includegraphics[width=0.40\textwidth,angle=-90]{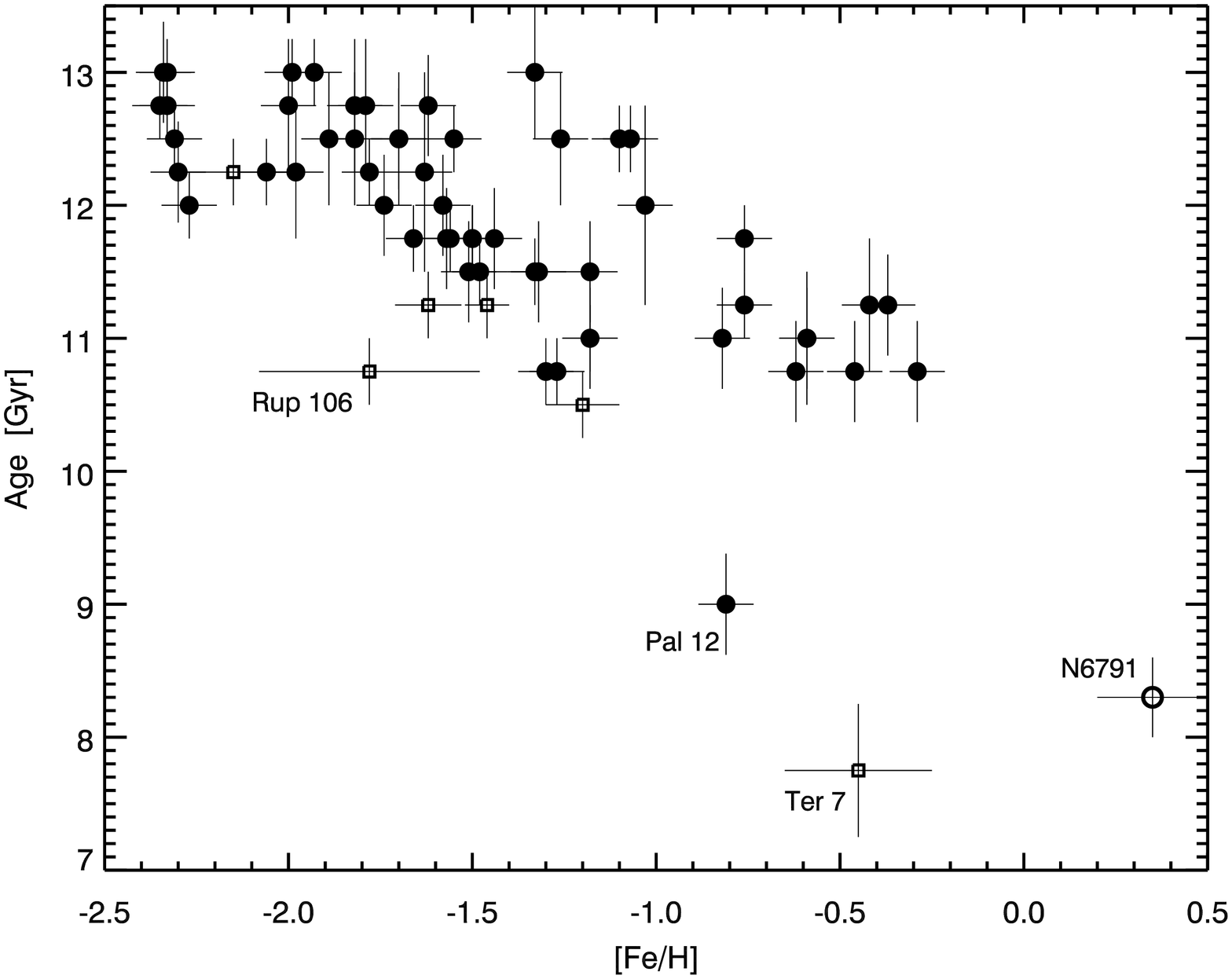}
\fi
\quad
\subfigure{\ifpdf
\includegraphics[width=90mm]{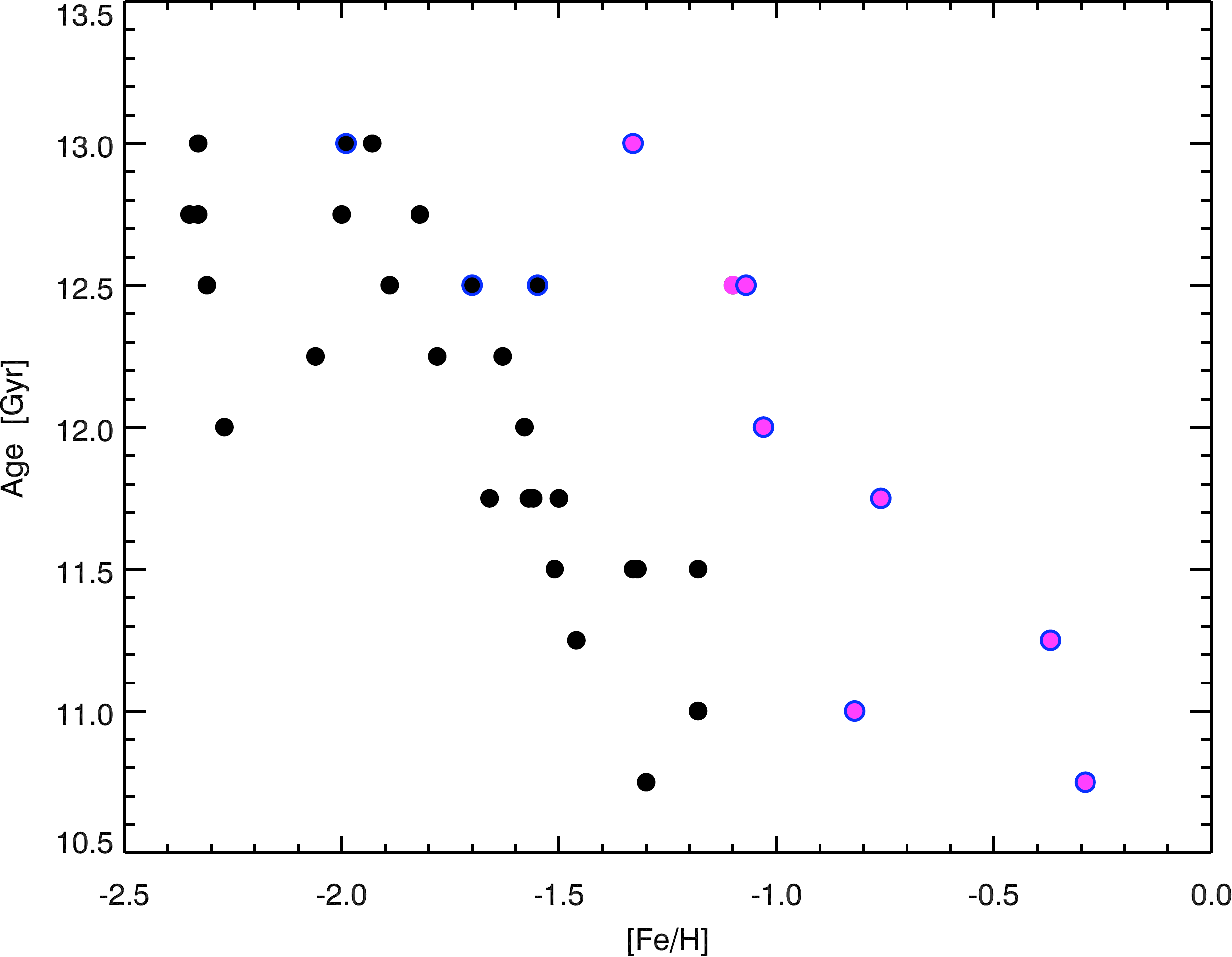}
\else
\includegraphics[width=0.40\textwidth,angle=-90]{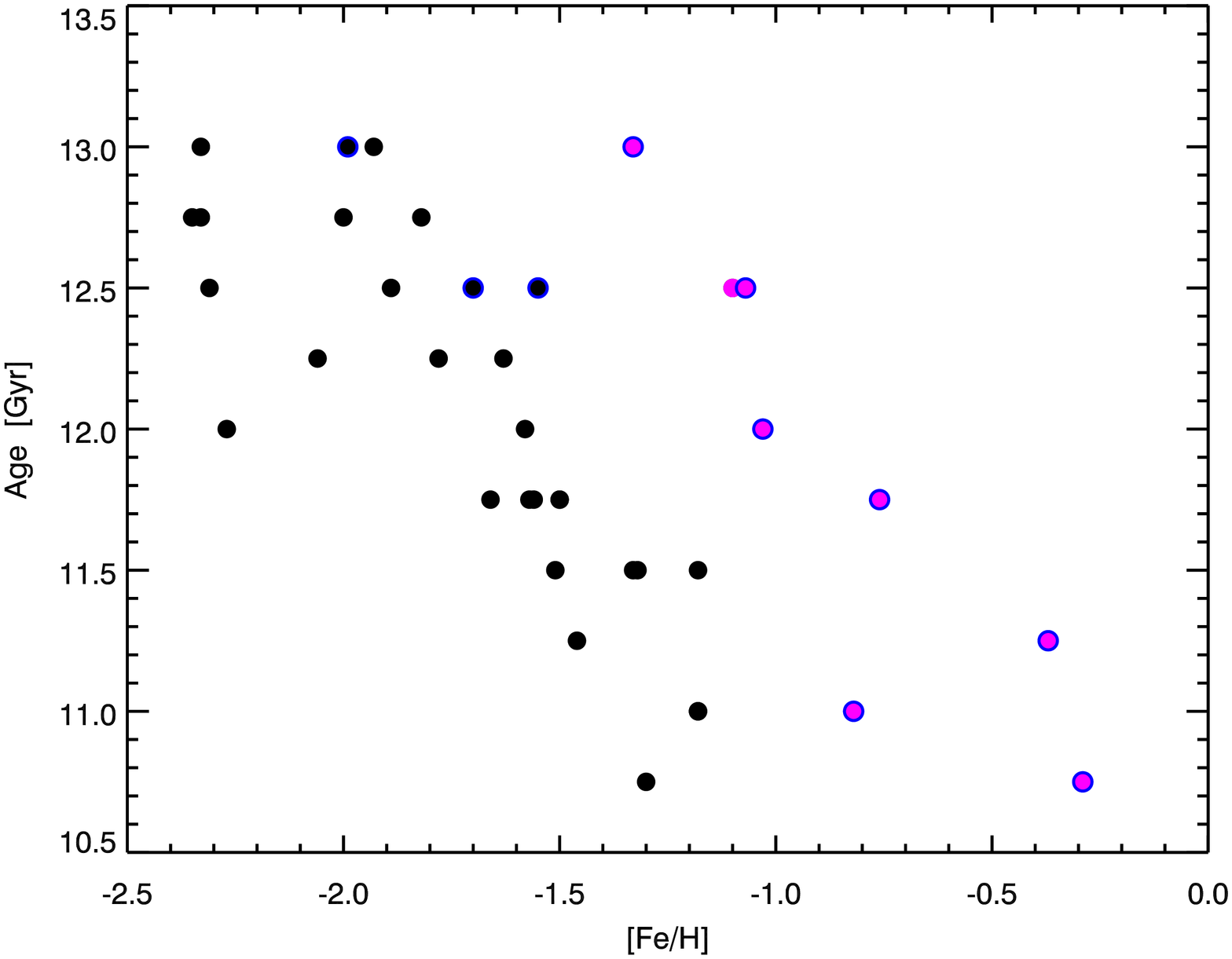}
\fi
 }}}
\caption{(\emph{Left panel}) Age versus metallicity for the Milky Way GCs from VandenBerg et
al.~(2013) shown as the black filled circles.  Six additional outer halo GCs
presented in this work are shown as the open squares, and the metal rich 
cluster NGC$\,$6791 from \protect\cite{Brogaard12} is plotted as the open circle.
The bifurcation of the
AMR into two ``arms'' is notable even down to the youngest ages.  
The right panel shows the AMR only for GCs which have phase space
information.  The points are colour coded by their probability of belonging to
the disk or halo as described in the text.  Blue circles indicate clusters
which are more likely associated with the MW disk, pink filled points the disk
clusters determined by eye simply by virtue of their metal rich offset in the AMR.
Notably, the disk clusters span the full range of ages, 
but occupy the more metal-rich arm of the two distinct AMR sequences.}
\label{fig:amrprob}
\end{figure*}

The indicated colour adjustments have been applied to the isochrones merely to
demonstrate that they accurately reproduce the photometry both just above and
just below the observed turnoffs and, therefore, that they have the same TO
luminosities as observed.  Such adjustments, regardless of whether they are due
to errors in the model temperatures or the colour transformations, in the
photometric zero-points, or in the assumed reddenings, have no impact on the
derived ages.  As found by V13, the red-giant branches (RGBs) of the best-fit
isochrones tend to lie slightly to the red of the cluster giants, which suggests
that the model temperatures or the adopted colour--$T_{\rm eff}$ relations or the
assumed metallicities are not quite right.  It is also possible that they are
due, in part, to errors in the luminosities of the ZAHB models, since fainter
ZAHBs would imply older ages and reduced TO-to-RGB colour differences.  However,
V13 derived ages of 13.0 Gyr for several of the GCs, and a significant upward
revision of their ages is unlikely given that the age of the universe is $13.77
\pm 0.06$ Gyr (\citealt{Bennett12}).  This is a moot point anyway since we are
much more interested in the differences in the ages of the GCs than in their
absolute ages.

A few comments should be made concerning Figure \ref{fig:ohalo}.  First, even
though Palomar 15 was included in the \cite{Dotter11} data set, we decided
against deriving an age for this system because it has a very blue HB along with
a high and presumably uncertain reddening ($E(B-V) = 0.387$, according to
\citealt{SFD98}), which make any fit of ZAHB models to the cluster HB population
very uncertain.  Second, with one exception (see the next paragraph), the
adopted [Fe/H] values were taken from the study by \citet{CG09} whenever
possible.  If the latter did not provide a metallicity estimate for a given
cluster, we opted to use the [Fe/H] value listed for it in the catalogue of
cluster properties by \cite{Harris10}.  Third, reddenings identical
to, or within 0.02 mag of, those given by Schlegel et al.~were assumed, though
higher or smaller values of $E(B-V)$ by $\approx 0.06$ mag were adopted in the
case of NGC$\,$6426 and Pyxis, respectively, in order to obtain reasonably
consistent interpretations of their CMDs.  As reported by V13, the isochrones
generally seem to predict TO colours that are too red by $\sim 0.01$--0.025 mag;
consequently, we arbitrarily set the $\delta$(colour) values for these two GCs
to be $-0.02$ mag and then adjusted their reddenings by the amounts needed to
achieve satisfactory matches to both the TO and HB stars.  Because the ZAHB
loci are nearly horizontal where the HB stars in these two clusters are located,
the impact of this approach on the derived ages is quite small ($\leq \pm 0.25$
Gyr).

Note that, whereas V13 did not attempt to determine an age for Terzan 7 because
the \citet{CG09} metallicity for it ([Fe/H] $= -0.12$) was outside of the
range for which stellar models were available, a closer examination leads us to
suspect that such a high value of [Fe/H] is unlikely to be correct.  According
to the \cite{SFD98} dust maps, the line-of-sight reddening in the
direction of Ter 7 is 0.100 mag, and the absorbing gas/dust must be mostly in
the foreground given that this GC has R$_{\rm G} = 15.6$ kpc \citep{Harris10}.
The intrinsic TO colour implied by this reddening suggests that the cluster
metallicity must be significantly less then $-0.12$.  In fact, several
spectroscopic studies (see \citealt[and references therein]{Pritzl05}) have
obtained [Fe/H] $\approx -0.6$, as compared with the value of $-0.32$ that is
given in the Harris catalogue.   We therefore chose to adopt [Fe/H] $= -0.45$ as 
a compromise of those determinations, and because this estimate also leads to
good agreement between the predicted and observed RGB slopes.  In addition,
we adopted [$\alpha$/Fe] $=0.0$ (e.g., \citealt{Taut04}), which appears to be
typical of other [Fe/H] $\ga -1$ GCs that are associated with the
Sagittarius dwarf galaxy (such as Pal 12, see \citealt{Cohen04}).

The other GCs considered in Figure \ref{fig:ohalo} were assumed to have normal
[$\alpha$/Fe] values for their metallicities (see V13) in view of the results
reported by, for instance, \citet{Smeckerhane02} and \citet{Mottini08}.  Although some
studies (e.g., \citealt{Brown97}, \citealt{Pritzl05}) found [$\alpha$/Fe] $\simeq
0.0$ for Ruprecht 106, they also reported [Fe/H] values that are 0.3--0.4 dex
higher than the value ([Fe/H] $= -1.78$) given by \citet{CG09}.  If the lower
metallicity is more accurate, as we have assumed, then Ruprecht 106 has
$\alpha$-element abundances that are not too different from other GCs that have
similar [Fe/H] values.  Further work is clearly needed to put the metallicities
of Ruprecht 106 and other GCs that have been subjected to limited spectroscopic
work, such as NGC$\,$6426 and Pyxis, on a much firmer footing.  However, such
uncertainties do not have major consequences for ages based on the
$\Delta\,V_{TO}^{HB}$ method because of its reduced sensitivity to the
abundances of the heavy elements (as already noted).  Indeed, the random
uncertainties of our age estimates are within $\sim \pm 0.25$--0.5 Gyr: it
is quite apparent from Figure \ref{fig:ohalo} that isochrones for our best
estimates of the cluster ages generally provide quite agreeable fits to the
observed CMDs.

Finally we include the intriguing metal-rich open cluster NGC$\,$6791, and adopt the
average metallicity and derived age from the detailed study of \citep{Brogaard12},
which used stellar evolutionary models and methodology consistent with our
age derivations of the other clusters.

\section{Phase Space Classification for the GCs}
In addition to the age and metallicity data discussed above, 
which trace a cluster's internal properties, it is also useful to 
consider GCs in the context of their host environment, in this case the MW.  
We therefore use GC phase space data from the compilation maintained by 
D. Casetti (and references provided therein; e.g., \citealt{Casetti07})\footnote{\url{http://www.astro.yale.edu/dana/gc.html}} to classify globular clusters
as either disk or halo populations using the probabilistic 
classification scheme described below. 

For each GC \emph{i} we can estimate the probability that it 
is associated with either the MW disk or halo as:
\begin{equation}
p(D,H{\mid} x_{i} v_{i}) \propto P(x_{i},v_{i}{\mid} D,H)P(D,H)
\end{equation}
where $P(x_{i},v_{i}{\mid}D,H)$ is the likelihood of observing the
phase-space coordinates $x_{i}, v_{i}$ ($x_{i} \equiv X_{i},Y_{i},Z_{i} , 
v_{i} \equiv U_{i},V_{i},W_{i }$) given the (known) kinematics 
of a particular sub-component and $P(D,H)$ is the prior probability 
for $GC_{i}$ to be associated with that component.  
In all cases we have assumed a uniform prior probability of membership. 
This analysis assumes that the kinematics of the MW GCs and 
field stars follow similar kinematic profiles - which holds in the 
general sense as far as differentiating halo- from disk-like orbits
for old populations. Dynamical friction, while altering the orbits of
accreted stars and GCs from dwarf galaxies, will not introduce
strong systematic uncertainties in this classification as the GC stars
are incorporated during cases of disruption,
contributing to the observed halo velocity distribution.

We describe the stellar density profiles of the Milky Way 
thin disk, thick disk, and halo using equations (22)--(24) 
of \cite{Juric08}, with the
best-fitting parameters for the various scale lengths and density normalizations
also taken from that work (their Table 10).  We assume that the stellar halo of the
MW extends to 150 kpc \citep{Deason12}, and following \cite{Pritzl05}, 
introduce a softening parameter $a = 20/\rho_{0}$ in order to keep the halo
density finite in the inner regions.  The velocity ellipsoids are taken to
be Gaussian with the mean values and the dispersions for the thin disk, thick
disk, and halo adopted from Table 3 of \cite{Venn04} and references therein.

There is substantial evidence (see the review paper by \citealt{RB13}) suggesting
 that the MW thick disk is not a distinct component (i.e., separate
from the thin disk), but rather represents a continuous extension of the thin
disk.  For example, \cite{Bovy12a, Bovy12b} showed that the structural (scale
heights and lengths) and dynamical properties of ``mono-abundance'' populations
in the MW disk smoothly change with chemical composition, therefore implying
that the MW has a single disk with a continuum of properties.  For this
reason, we do not make a distinction between thin- and thick-disk systems, but
simply classify them as members of the disk distribution in our calculations.
Taking, $p_{disk} = p_{thin} + p_{thick}$, we assume the probability of a
cluster belonging to either the MW disk or halo as:
\begin{align}\nonumber
{\rm log}( \it{p_{disk}}/\it{p_{halo}})&\gid 0:~  \mbox{ (for the disk)}\\
&\lid 0:~  \mbox{ (for the halo).}
\end{align}

\begin{figure*}
\centering
\mbox{\subfigure{\ifpdf
\includegraphics[width=90mm]{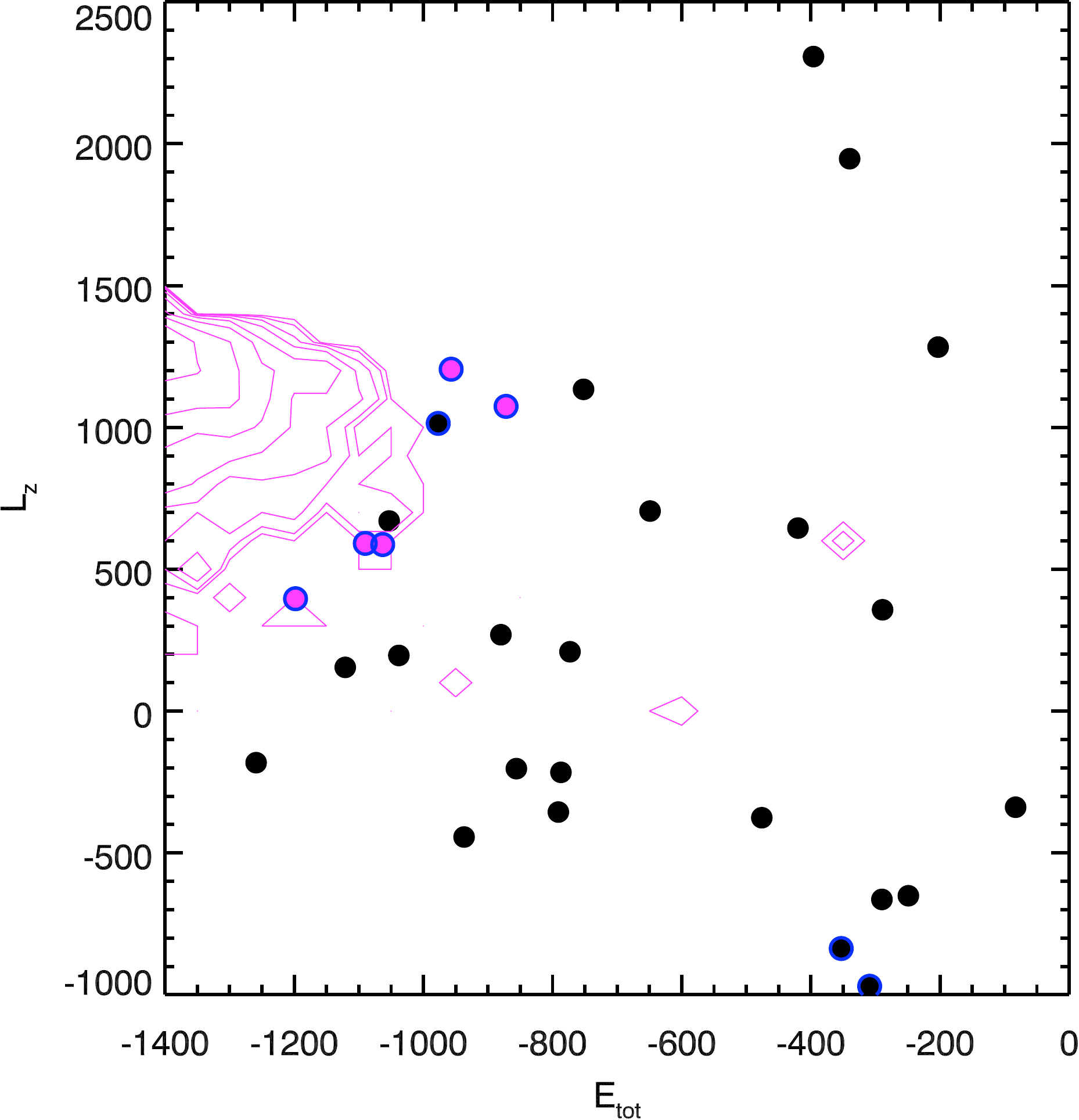}
\else
\includegraphics[width=90mm,angle=-90]{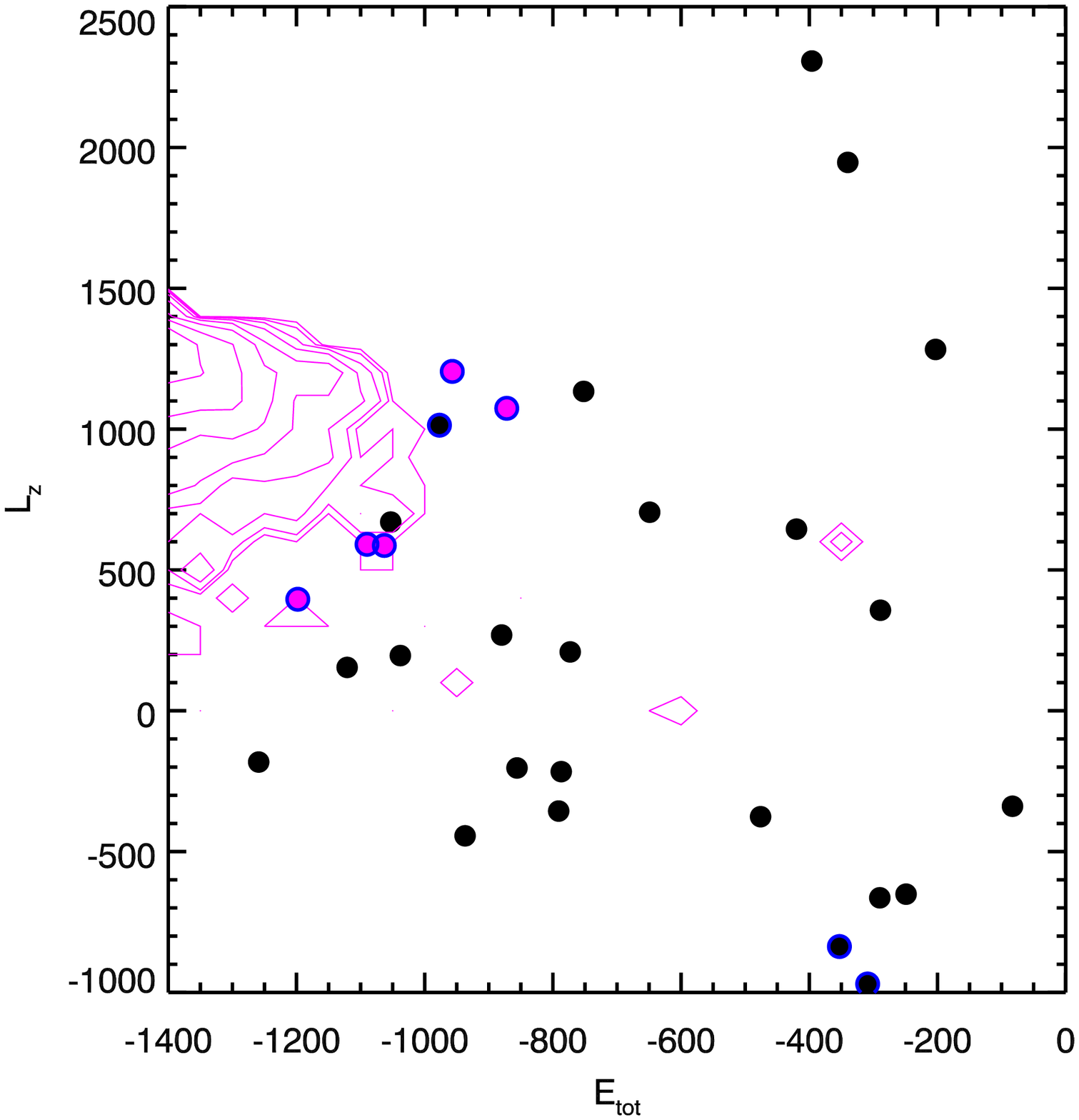}
\fi
\quad
\subfigure{\ifpdf
\includegraphics[width=90mm]{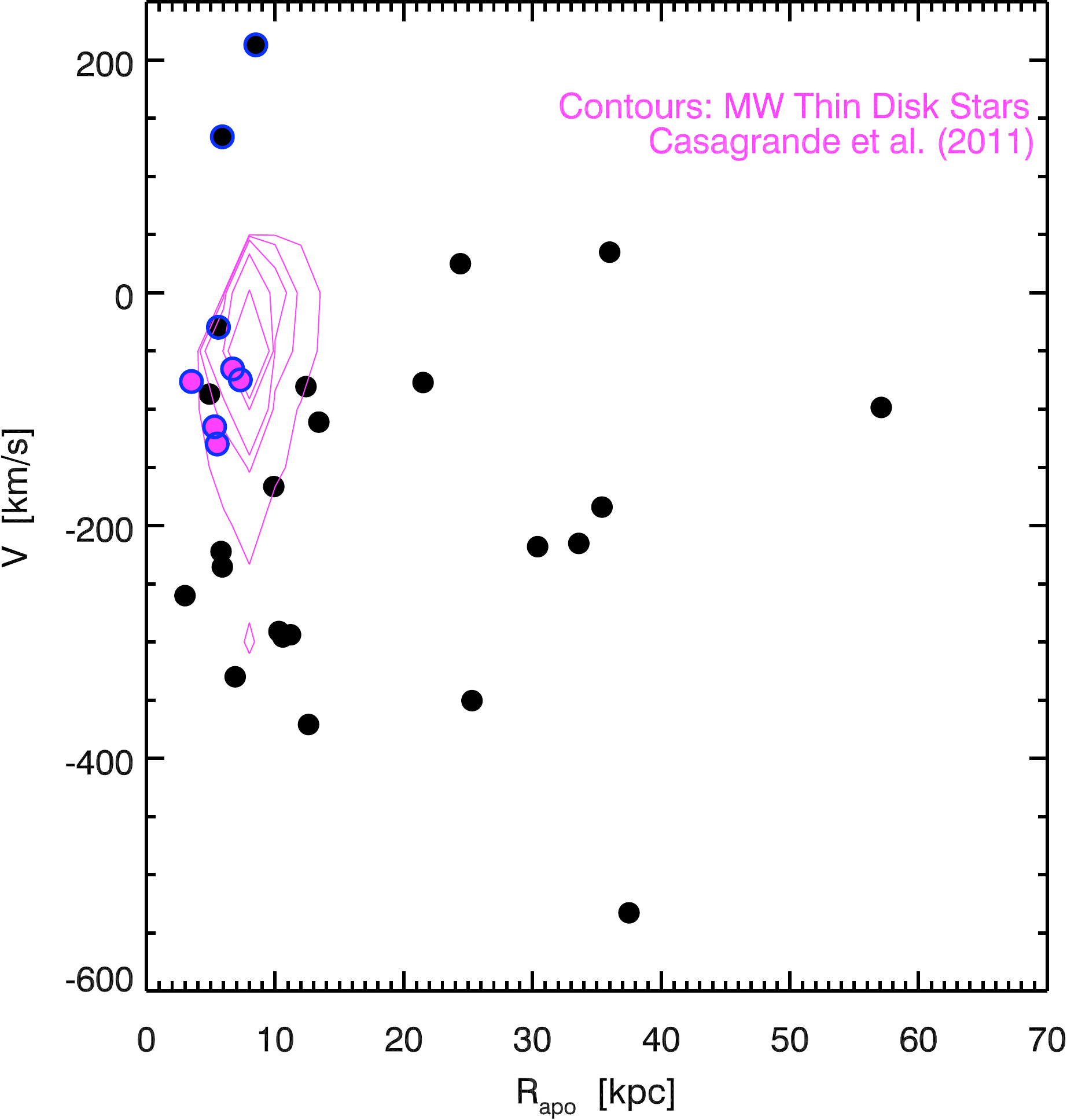}
\else
\includegraphics[width=90mm,angle=-90]{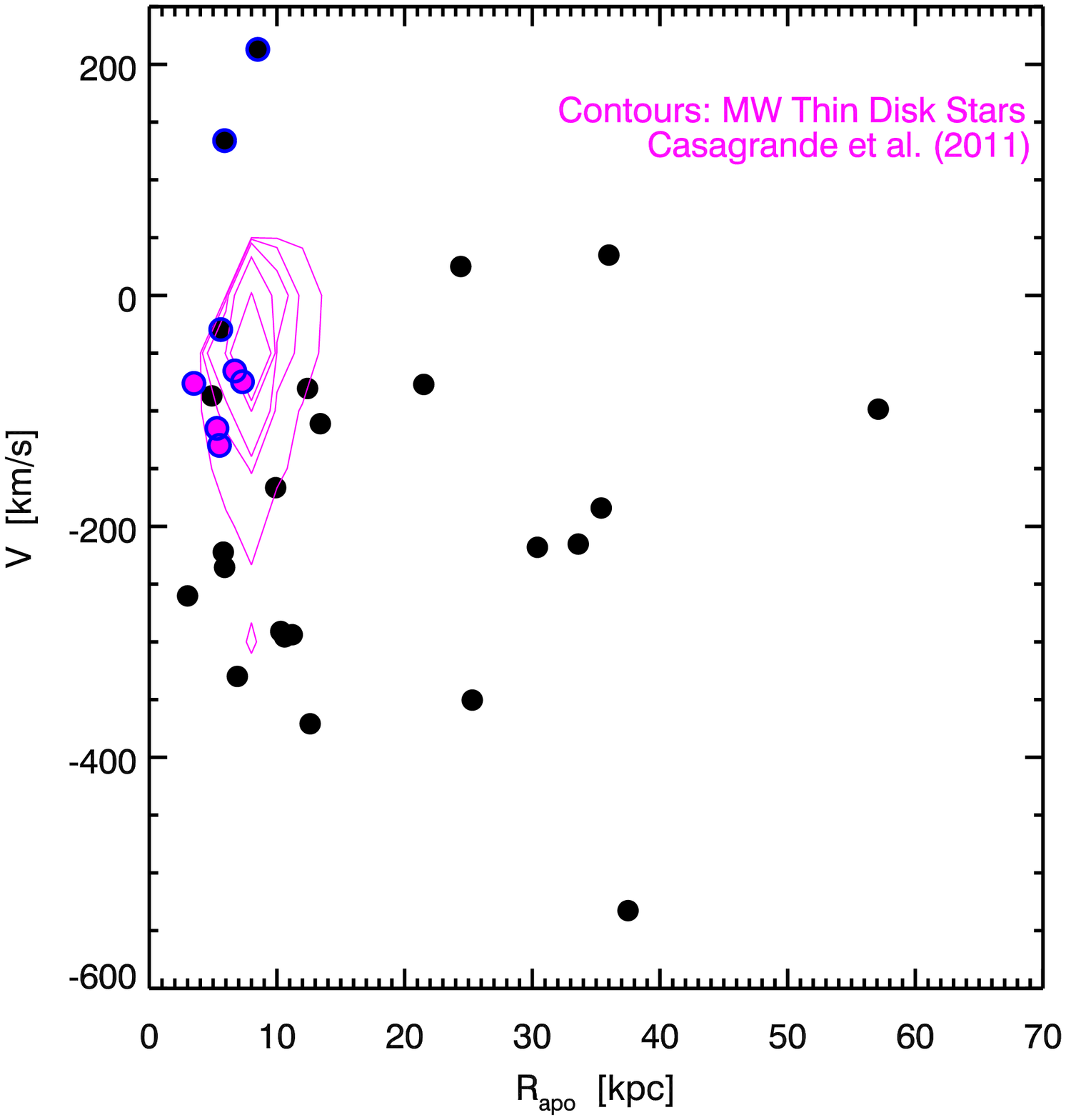}
\fi
 }}}
\caption{Common orbital properties for the clusters associated
with the Milky Way disk (pink dots).  The left panel plots the angular momentum
versus total orbital energy, whereas the right panel plots the Galactocentric
$V$ velocity versus apocentric radius of their orbit.  The disk GCs cluster
together and also tend to overlap with the MW disk stars
(\emph{pink contours}) from \protect\cite{Casagrande11}.}
\label{fig:orb2pan}
\end{figure*}

Moreover, we do not separate the halo populations as young (YH) or old (OH) as
in the study by \cite{Mackey05}.  The classifications in that work are based on
the horizontal branch type as a function of metallicity, assuming that age is
the main second parameter controlling the HB colour; however, the influence of
possible helium enhancements on that metric complicates such divisions (see e.g.,
\citealt{Gratton10}).  Regardless, the halo clusters are canonically thought to
have been accreted --- see, e.g., \citet{Mackey13} for evidence in support of an
accretion origin for 80\% of the GC population in the outer halo of M31, as well
as recent work by \cite{Keller12}, which reports similar findings for the YH
clusters in the MW (see also \citealt{Muratov10}). Hence, for the present
exercise, it is reasonable to treat all halo clusters as a single class.

\section{The Split Milky Way GC Age-Metallicity Relation}
The upper panel of Figure \ref{fig:amrprob} shows the AMR for all 61 Milky Way
GCs in our sample.  A split in the GC AMR is evident, with an offset metal-rich
sequence running from approximately [Fe/H]$\sim -1.5$ and 13.0 Gyr down to
[Fe/H] $\sim -0.4$ and 10.75 Gyr.  The right panel plots as blue
circles those clusters that have been identified as members of the disk from
our Bayesian classification. The
remaining black dots that populate the metal-poor branch of the AMR are
associated with the halo. 
As the kinematic classification is dependent on the assumed density
profile and velocity ellipsoids for the MW components, we also show with
pink circles, those clusters that one would pick ``by eye'' as belonging
to the offset AMR sequence.  The kinematically classified disk clusters include
all of those that one would select ``by eye'' as members of the offset
sequence, giving us confidence that the classification scheme is appropriate.

Of the kinematically classified disk clusters which lay on the 
metal-poor branch of the AMR (blue circles around black dots), all have been listed
in literature studies as being halo clusters from high mass progenitors, 
and are noted to have orbits which have likely been strongly affected by 
dynamical friction.  In fact \cite{Dinescu99} and \cite{Dinescu13}
 explicitly list NGC 6656, 6752,
6397, and 6254 as clusters which are likely halo members but whose orbits have been
made more disk-like through orbital decay.  The sole cluster in the disk sequence
which is classified as a halo member is NGC 6723 (confirming the conclusions of 
\citealt{Dinescu03}), although interestingly this cluster has kinematics and a location
that confine it to the MW bulge.  In conclusion, the most
secure phase space classifications are completely consistent with a 
simple division of the AMR into two branches ``by eye''.

\begin{figure*}
\centering
\mbox{\subfigure{\ifpdf
\includegraphics[width=96mm]{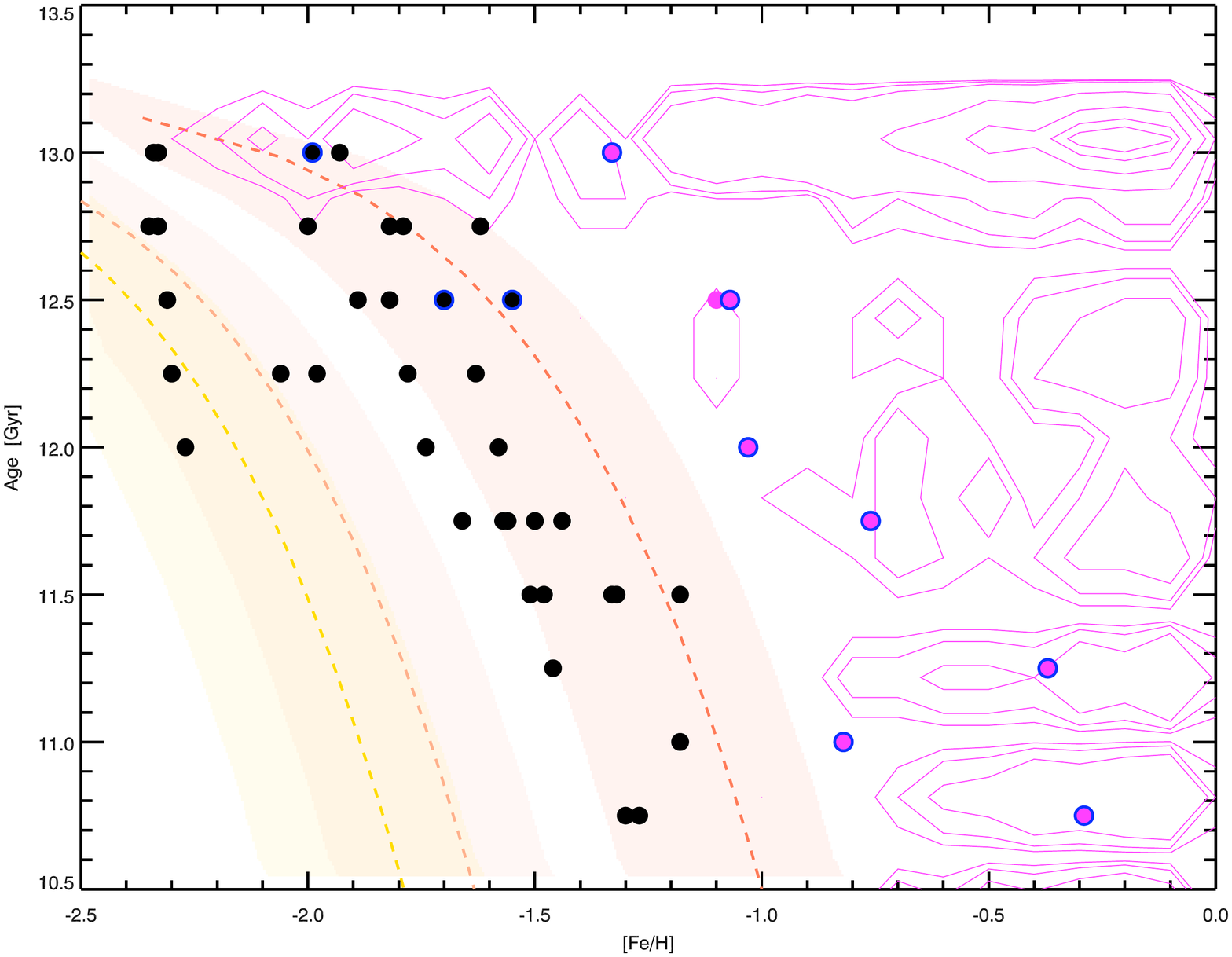}
\else
\includegraphics[width=0.407\textwidth,angle=-90]{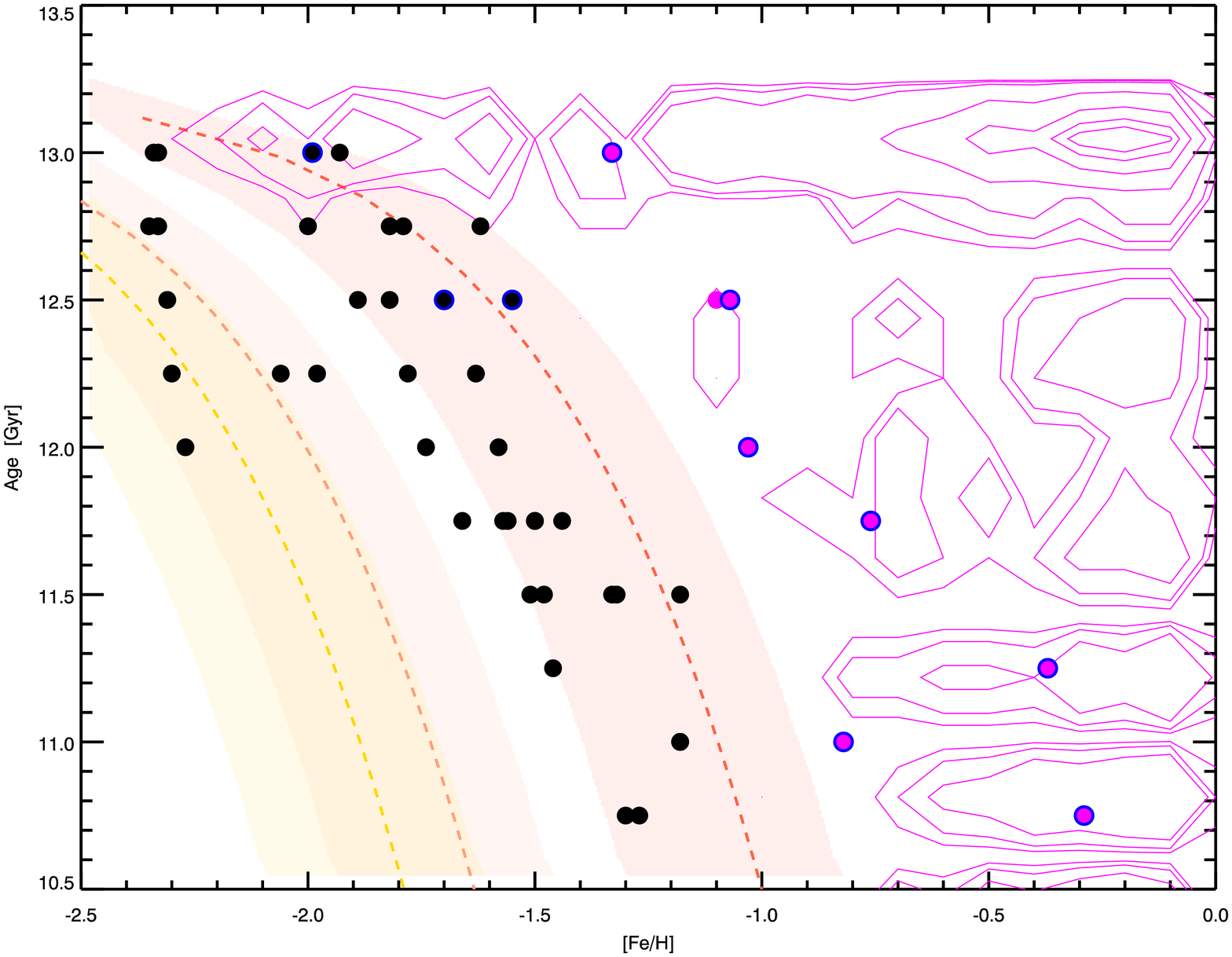}
\fi
\quad
\subfigure{\ifpdf
\includegraphics[width=90mm]{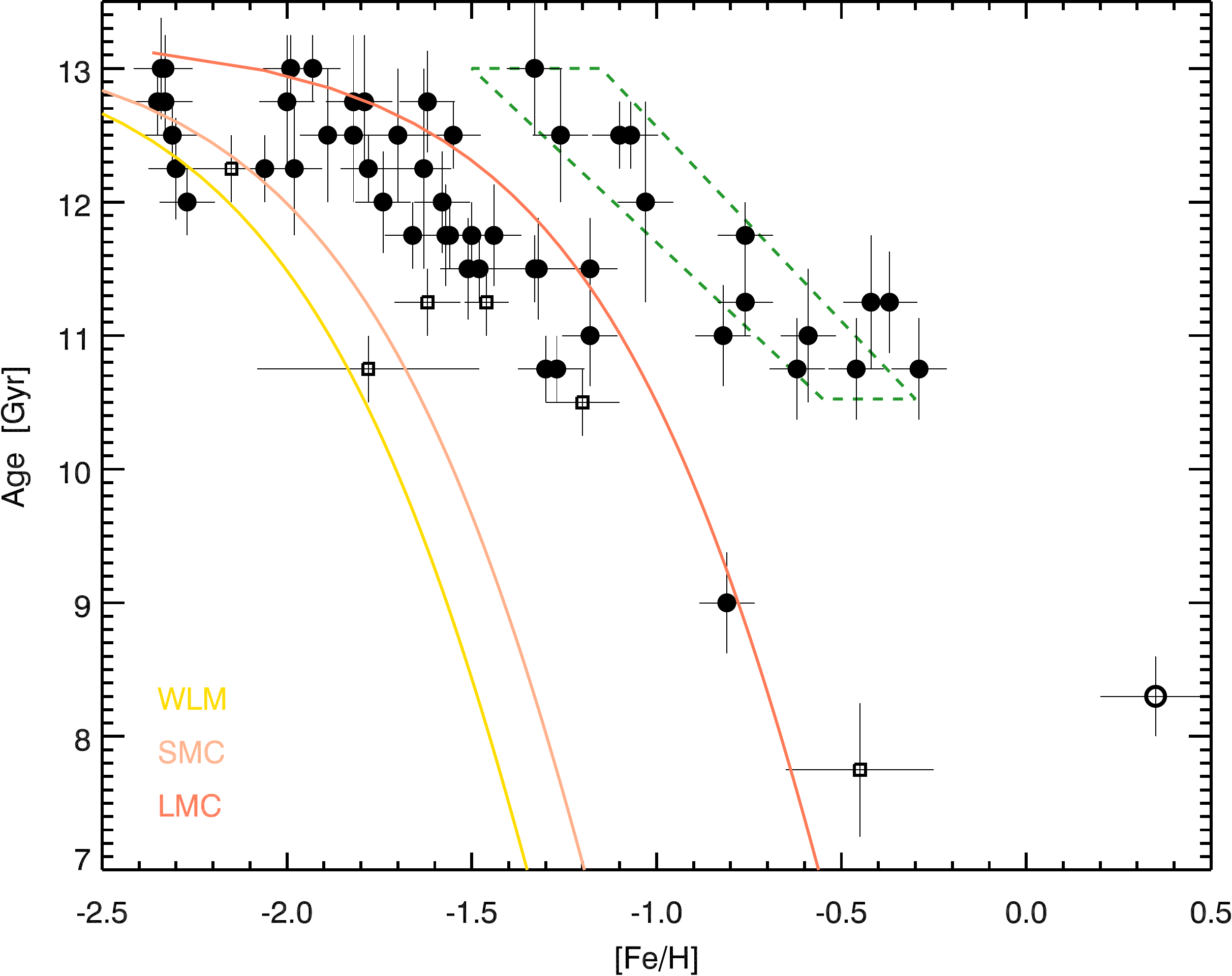}
\else
\includegraphics[width=0.39\textwidth,angle=-90]{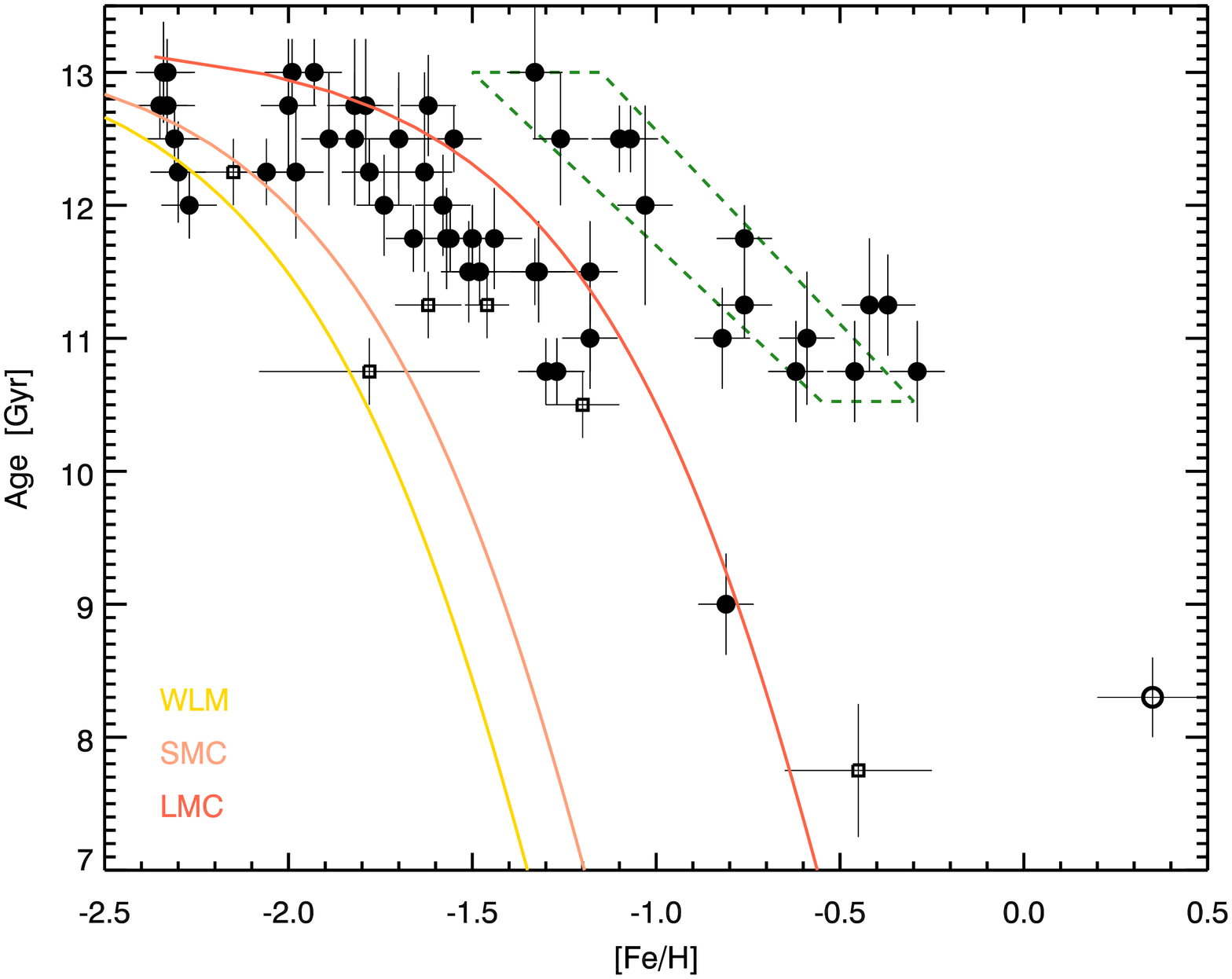}
\fi
 }}}
\caption{AMR of the MW GCs divided into the halo clusters and disk clusters,
as determined from their orbital and phase space characteristics.  Overlaid
as the pink contours is the AMR from MW thin disk stars \citep{Casagrande11}
which show good agreement with the disk GCs.  The AMRs for dwarf galaxies of
$M_{*} \sim 10^{7-9}$ from \protect\cite{Leaman13} and references therein are
also overlaid, and the close
agreement with the halo GCs suggest that they likely formed from dwarf galaxies
of such masses and were accreted during the formation of the MW halo.
Right panel shows the full range of age and metallicities, with the six new
halo clusters shown as black squares, as well as a predictions for the AMR of the
MW bulge GCs shown as the dashed green region (see \S 7.3). }
\label{fig:gcamrs}
\end{figure*}

It is somewhat reassuring that Pal 12 and Terzan 7 lie on the extension of the Halo AMR 
given that these clusters are thought to be associated with the accreted Sagittarius dSph.
\cite{Law10} identify M54, Terzan 8 and Arp 2 as GCs with a high probability
of association with Sagittarius, while NGC 5053, Pal 12 and Terzan 7 are moderately likely to be
associated.  However the present MW AMR unfortunately offers little in the way of 
constraints on their membership, as halo GCs which are certainly not associated
with Sagittarius also overlap with them in this parameter space. 
Therefore such membership studies 
will still be best probed through simulations of the tidal disruption of 
Sagittarius.

Figure \ref{fig:orb2pan} plots the total orbital energy and Z-component of
the orbital angular momentum, along with the Galactocentric velocity ($V$) and
orbital apocentric distance ($R_{apo}$) of the clusters.  The disk clusters not
only show common orbital properties, but also overlap with the data for MW thin
disk stars (the density contours which are shown as magenta lines): the latter
were derived from the updated analysis \citep{Casagrande11} of the 
Geneva-Copenhagen Survey \citep{Nordstrom04}.

Having identified those clusters which are most likely associated with the disk
of the MW, an obvious interpretation of the metal-rich arm of the split AMR is that
it contains GCs that formed in-situ in the disk.  It should be noted that the
disk clusters are not simply the most metal-rich clusters, but the most
metal-rich ones at any given age --- and that they span the full range of ages
encompassed by the halo clusters.  The slopes of the AMR sequences are steep 
enough that it is not possible to make a simple cut 
at constant [Fe/H]$\ga -1.5$ and have a ``clean'' sample of disk clusters.

If we assume that GCs metallicities trace the metallicities of their hosts 
when the bulk of stars formed, then a plausible interpretation of the offset 
between the metal-rich disk and the
metal-poor halo AMRs follows from consideration of the galactic
mass-metallicity relation (MMR).  The offset in the GC AMR is approximately
$0.6$ dex in metallicity, which, given the slope of the MMR\footnote{We note
that the slope and differential shape of both the gaseous and stellar MMR in 
the mass range of interest are nearly identical \citep{Lee08,Kirby11}} 
\citep{Tremonti04,Gallazzi05,Lee06,Kirby11}, translates into a difference in stellar mass
of approximately $\Delta\log\,M_{*} \sim 2$ dex.  
Since the MW disk has a mass of $\sim (3 \pm 1) \times
10^{10} \msol$ \cite{Robin04,McMillan11}, 
this implies that the halo GCs are described by an AMR that is
representative of a galaxy with a stellar mass of a few $\times 10^{7-8} 
\msol$.\footnote{ We note that 
the \textit{relative difference} in
specific SFR \citep{Karim11} and metallicity \citep{Zahid13} 
between two galaxies of different mass stays roughly constant with
time (back to redshift 3) as they evolve --- at least for masses similar to the 
MW and the LMC.}  This
suggests that the halo GCs would have formed in dwarf galaxies comparable to
the SMC, WLM, or even the LMC and Sagittarius.  

This is empirically illustrated in Figure \ref{fig:gcamrs},
where the AMRs for the MW thin disk, as well as those that have been derived for
three dwarf galaxies are overlaid\footnote{We do not show the Sagittarius AMR
due to the extreme difficulty in selecting clean, representative samples of RGB stars
in this object, however the AMR  
given by \cite{Law10} follows a similar shape and lies between
that of the LMC and SMC.} on the GC AMRs.  The dwarf galaxy AMRs, which come from
spectroscopic measurements of individual RGB stars, were compiled and presented by
\cite{Leaman13} and the data originally analyzed in the studies of
\cite{Cole05,Pompeia08,Carrera08,Carrera08b,Leaman09,Parisi10,Leaman12b}.  

The disk clusters tend to coincide 
with the metal poor edge of the MW thin disk AMR.  
This is likely because of the well known impact of
radial migration \citep{Sellwood02,Roskar08} in 
scattering the disk AMR to higher metallicities at
a given age --- as well as the fact that we are considering clusters from the entire disk
and some may be more closely linked to a metal poor ``thick'' disk component
than the pure thin disk AMR.  In addition the \citep{Casagrande11} ages were 
derived using stellar evolutionary models without diffusion, though this effect
is likely below the level of the previous two systematics.  

Importantly, as the disk AMR from 
\cite{Casagrande11} consists of primarily bright solar neighbourhood stars,
it may provide only a partial view of the total disk AMR
if the MW disk exhibits a strong metallicity gradient.  A Complicating factor is that radial
migration also shifts the observed AMR away from the ``true'' disk AMR.  
The interplay between these two
systematics on the observed AMR is nearly impossible to quantify given the
unknown evolution of the MW metallicity gradient, and when coupled with the $1-3$ Gyr
errors on the disk star ages, highlights the fact that we simply do not have an
unbiased MW disk AMR with which to compare the disk GC's AMR.  Nevertheless
the disk GCs are clearly separated and distinct from the halo GCs and even 
the most massive dwarf galaxy AMRs.

In this context where the age and metallicity of a cluster are intimately
linked to the mass of its birth galaxy, NGC$\,$6791 does not appear so unusual.
This cluster has been extensively studied due to the apparent contradiction
of having an old age, but high metallicity (for an open cluster), 
or an age much younger than is characteristic of metal-rich GCs.  However it is clear from
Figure \ref{fig:gcamrs} that it follows an extension of the AMR for the MW disk 
(or bulge; see $\S 7.3$), and in that regard may be completely consistent
with expectations for clusters born out of a high mass progenitor environment. 

In summary, the disk clusters exhibit an excellent empirical
agreement with the AMR of the
most metal poor MW disk stars from the 
Geneva Copenhagen Survey; while the AMR of the
halo GCs spans the range shown by the three dwarf galaxies.

\section{Estimating the Accretion History of the Milky Way from its GC System}
The halo clusters that form the metal-poor branch of the MW GC AMR are
well fit by the AMRs of several Local Group dwarfs.  However, this does not
provide any explicit information on the relative number of those systems that
may have been accreted by the MW.  Several studies have attempted to place
constraints on the number of accreted clusters (and their percentage of the
halo mass), often using the variation of the structural or HB properties of the
clusters with their position in the Galaxy to infer which ones may have been
accreted (e.g., \citealt{Mackey05,Forbes10})

Having the luxury of the split AMR and orbital data that enable us to classify
the GCs into in-situ disk and halo systems, we may study this question further
given the explicit link between the halo clusters and the mass(es) of their
progenitor host (dwarf) galaxies.  In Figure \ref{fig:gcamrs} there are 
$\sim 4-6$ clusters which
are consistent with originating from a low mass WLM sized dwarf, while a 
great number more ($\sim 25-30$) are associated with a higher mass LMC-like
progenitor.  Is this consistent with expectations for the merger history of
the MW --- and with the expected number of GCs which might be accreted during
the merger of dwarf galaxies with our Galaxy?

Answering this question requires comparing the observed number of halo 
clusters from the split AMR and the total stellar mass of
the MW stellar halo ($\sim 1 \pm 0.4 \times 10^{9} \msol$; 
\citealt{Morrison93,Bell08,Deason11}) to simulations of the 
merging history of the MW to check whether the GC population and stellar mass of 
the MW halo could have been accreted self-consistently.  The new observational
constraints from the mass-dependent splitting of the AMR in Figure 
\ref{fig:gcamrs} provide a first step in understanding how many
dwarfs of a given mass merged to build up the stellar
halo and the GC population of the MW.

To do this requires an estimate of the number of dwarfs of a given mass which
merged with a MW-sized halo, as well as an expectation for the GC specific
frequency $S_{N}$\footnote{Defined as the total number of GCs hosted by a
galaxy, normalized to its luminosity: $S_{N} = N_{GC}10^{0.4(M_{V} + 15)}$} for the
dwarfs that were accreted.  The differential number of subhalos of dark matter
mass $m$ that a larger galaxy of mass $M$ accretes has been shown to be robust
function of the total mass of the primary galaxy and its redshift, as given
explicitly by \cite{Giocoli08} as:
\begin{equation}
N_{merged}(m) = A \int m^{-1.8} dm
\end{equation}
The normalizing constant assumes that 25\% of the stellar mass of the primary
galaxy is accreted from smaller galaxies that have a mass fraction between
$10^{-5} \leq \frac{m}{M} \leq 10^{-2}$.  This gives the total number of merged
subhalos as a function of their dark matter mass.  We also explore an 
alternative formulation of the subhalo mass function from the 
high resolution $n$-body Aquarius Simulation of MW-sized halos; 
specifically, equation (7) of \cite{BK10}.

To associate a stellar mass $m_{*}$ with every subhalo of dark matter mass $m$,
we use the stellar-to-halo mass relation (SHMR) from \cite{Leauthaud12}, which
has been robustly computed from an analysis of the observed weak lensing and
spatial correlations in galaxy populations: this provides a measure of the
stellar mass of a galaxy with a given dark matter halo.  As in other works
\citep{Guo10} the SHMR in the linear low mass regime is extrapolated for masses
below $M_{*} \leq 10^{8}$.  Together with equation
(3) above, it is possible to compute the number of dwarfs of a given stellar
mass which merge with the MW (assuming that its total dynamical mass is $\sim 8
\times 10^{11} \msol$; \citealt{Veraciro13}).  This procedure thus
enables us to track the buildup of the stellar halo.  

\begin{figure}
\begin{center}
\ifpdf
\includegraphics[width=0.49\textwidth]{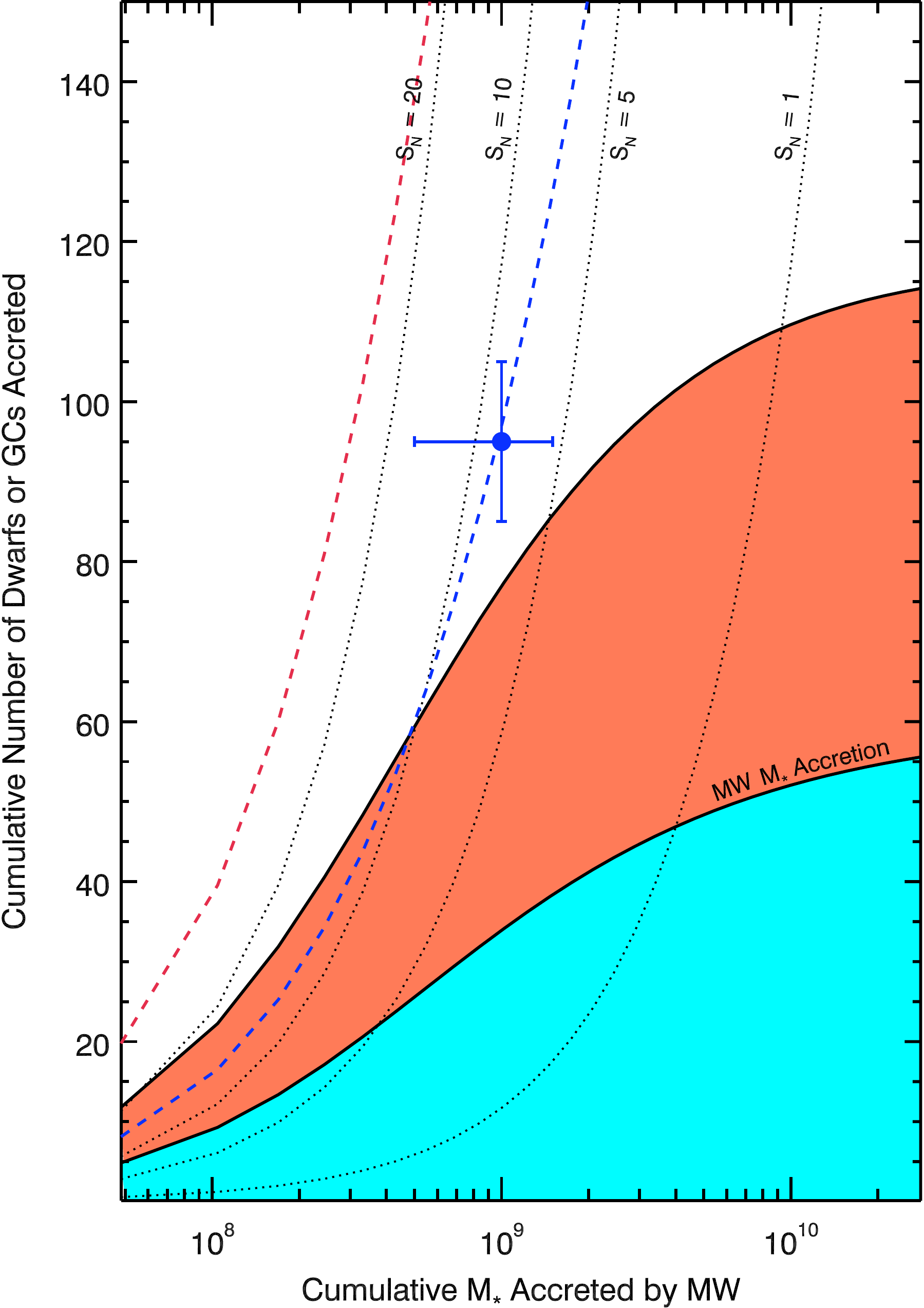}
\else
\includegraphics[width=0.49\textwidth]{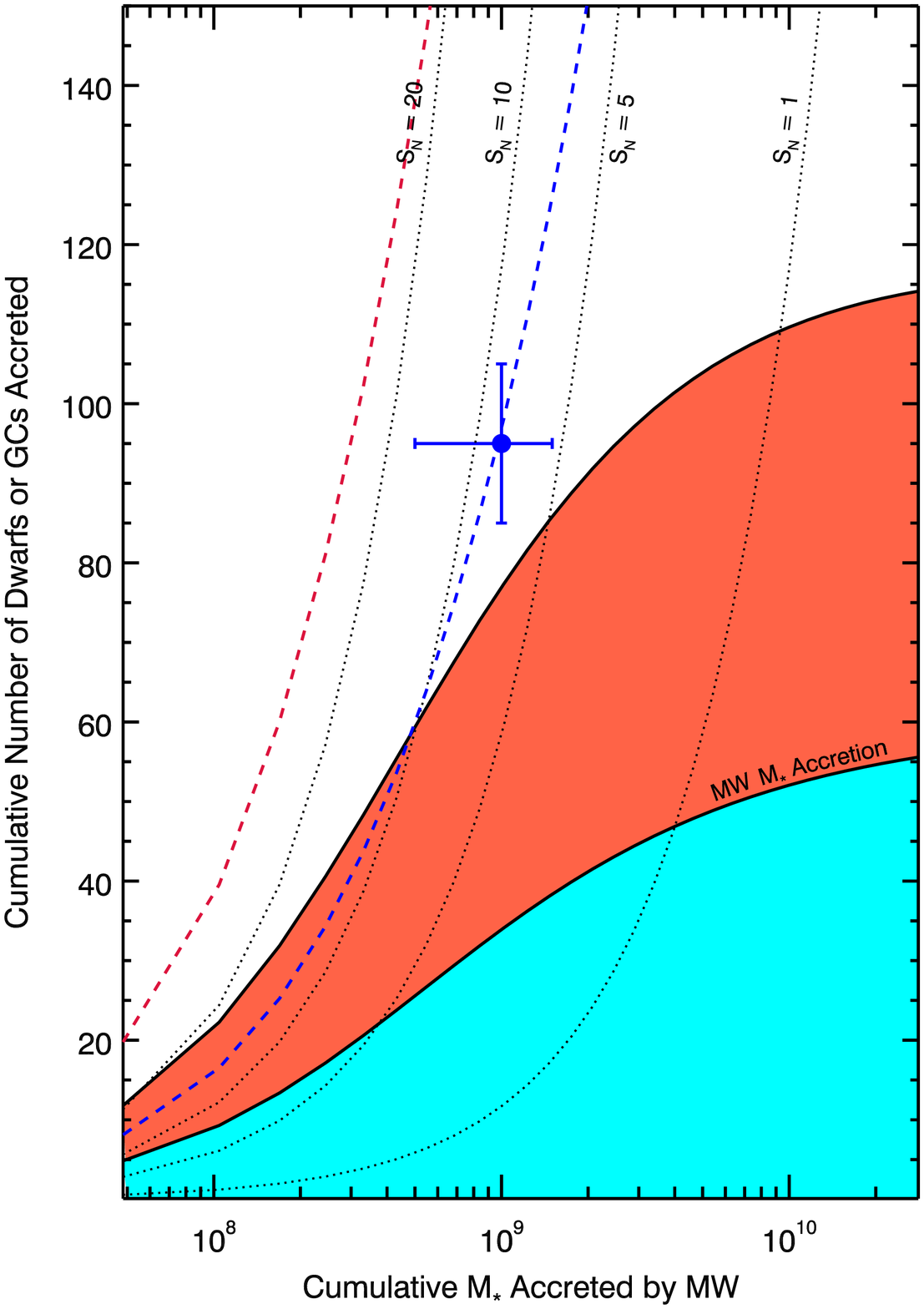}
\fi
\caption{Cumulative stellar mass accreted by the Milky Way in assembly of its
stellar halo, plotted versus the cumulative number of subhalos (dwarf galaxies)
or GCs merged.  Two subhalo mass functions are shown, one from \protect\cite{Giocoli08}
(\emph{solid black line bounding blue region}), and one based on the merger
history of the MW-sized halos in the Aquarius Simulations \protect\citep{BK10}
(\emph{black line bounding orange region}).  These solid lines show the cumulative
number of accreted dwarfs by the MW.  For various specific frequencies
of GCs (dotted black lines), it is possible to show the cumulative number
of GCs that were accreted in the same process.  Observed number of halo GCs in the MW
and stellar halo mass for the MW is shown as the blue point.  While a fixed
average $S_{N} \sim 7$ for all the merging dwarfs is possible, the dashed blue
line shows that adopting a $S_{N}$ which varies with mass as seen in
observations of galaxies by \protect\cite{Peng08}, reproduces the MW system without any
fine tuning.  Red dashed line shows the same \protect\cite{Peng08} $S_{N}$ relation, but
assuming the merger history of the Aquarius Simulations.}
\label{fig:sngc2}
\end{center}
\end{figure}

The crucial question of how many GCs are contributed by an accreted dwarf
requires an assumption about the specific frequency of GCs for each dwarf.
Values of $S_{N}$ typically range between 0.5 and 10, with some outliers, as in
the case of M$\,$87 and Fornax which have $S_{N}$ values approaching 20
\citep{Mateo98,Peng08}.  There is also evidence that $S_{N}$ itself varies
systematically with the luminosity or mass of a galaxy (i.e., \citealt{Peng08}) -
which may be linked to the galaxy SHMR (see also, \citealt{Spitler09}).
It is necessary to examine the implications of assuming different values 
for the specific frequencies, as we have no {\it a priori} knowledge 
of what the values of $S_{N}$ were for the dwarfs that were accreted by the MW.

To estimate how many GCs will be accreted into the MW along with their host
dwarf galaxies, we compute the expected number of accreted dwarfs of stellar 
mass $m_{*}$ as given by equation (3) and the SHMR, 
and multiply the result of that calculation by the number of
GCs belonging to each dwarf of a given mass (which depends on a choice for
$S_{N}$). The cumulative total of this product
provides estimates of both the amount of stellar mass and the number of GCs 
that are contributed by accreting dwarf galaxies to the MW stellar halo.
We repeat this for various fixed values of $S_{N}$ as well as the 
mass dependent form of $S_{N}$ from \cite{Peng08} which is extrapolated
 to lower masses using the SHMR of \cite{Leauthaud12},
 as has been similarly done in \cite{Spitler09}.

Figure \ref{fig:sngc2} plots the \emph{cumulative} stellar mass accreted by
the MW on the x-axis, versus the number of merged subhalos or GCs on the y-axis.
The solid black lines show the cumulative mass growth of the MW stellar halo as
the number of mergers increases for the subhalo mass functions of equation (3)
and the relations in \cite{BK10}.  The dotted lines indicate, for different specific
frequencies, how many GCs are accreted by the MW as its stellar halo is
assembled.

The blue dot indicates the observed values of the MW stellar halo mass, and the
number of halo GCs.  We have assumed that the ratio of disk to halo GCs seen in
our sample ($N_{disk}/N_{halo} \sim 20$--30\%) is representative of the total
GC population of the MW.  On the assumption that the Galaxy has a total of
$N_{tot}$ = 150 GCs, of which 20 conceivably belong to the bulge, we obtain
$N_{GC} = 95-105$ (the blue cross value) for our estimate of the current number
of halo clusters.

At the observed stellar halo mass of $\sim 10^{9} \msol$,
Figure \ref{fig:sngc2} indicates that the MW may have accreted $\sim 25-35$ dwarf
galaxies (as indicated by the intersection of a vertical line at this mass with
with the relevant solid black curve).  The Aquarius simulations predict a
similar form for the growth of the mass that is accreted by the MW, but suggest
that $\sim 2$ times as many dwarfs were merged in order to add a corresponding
amount of mass. These dwarfs could easily contribute enough GCs to account for
the entire population of the MW halo GCs, if they had, on average, a specific GC
frequency $S_{N}\sim 7$.  The dashed lines indicate how many GCs would be
produced as a function of the accreted stellar mass if we used the mass
dependent $S_{N}$ scalings from \cite{Peng08} for each of the two subhalo mass
functions.  The blue line shows excellent agreement with the MW values while
the red line, which assumes the relations from the Aquarius Simulation, would be
consistent with lower values for the MW stellar halo mass (or
higher total numbers of halo GCs).  This could easily be the case if some 
fraction of the MW's stellar halo was formed in situ, and/or additional outer halo
GCs are discovered. 

This self-consistent check on the accretion history of the MW is jointly 
constrained by the number of observed GCs in the halo, and its stellar mass.
To first order this approach seems to offer useful constraints, nevertheless there
are some obvious caveats; e.g., what fraction of stellar mass is fully
accreted, and how many GCs survive?  However we note that the observational
constraints and relative fraction of accreted stellar mass are in good
agreement with the recent results of \cite{Cooper13} which tracked the
assembly of stellar mass in high resolution n-body and semi-analytic simulations.
In particular that work showed that MW sized galaxies could accrete $1-50$\% of
their total (including disk and halo) stellar mass, and this was enough to 
\emph{fully} assemble the stellar halo of the MW through mergers.

\begin{figure}
\begin{center}
\ifpdf
\includegraphics[width=0.49\textwidth]{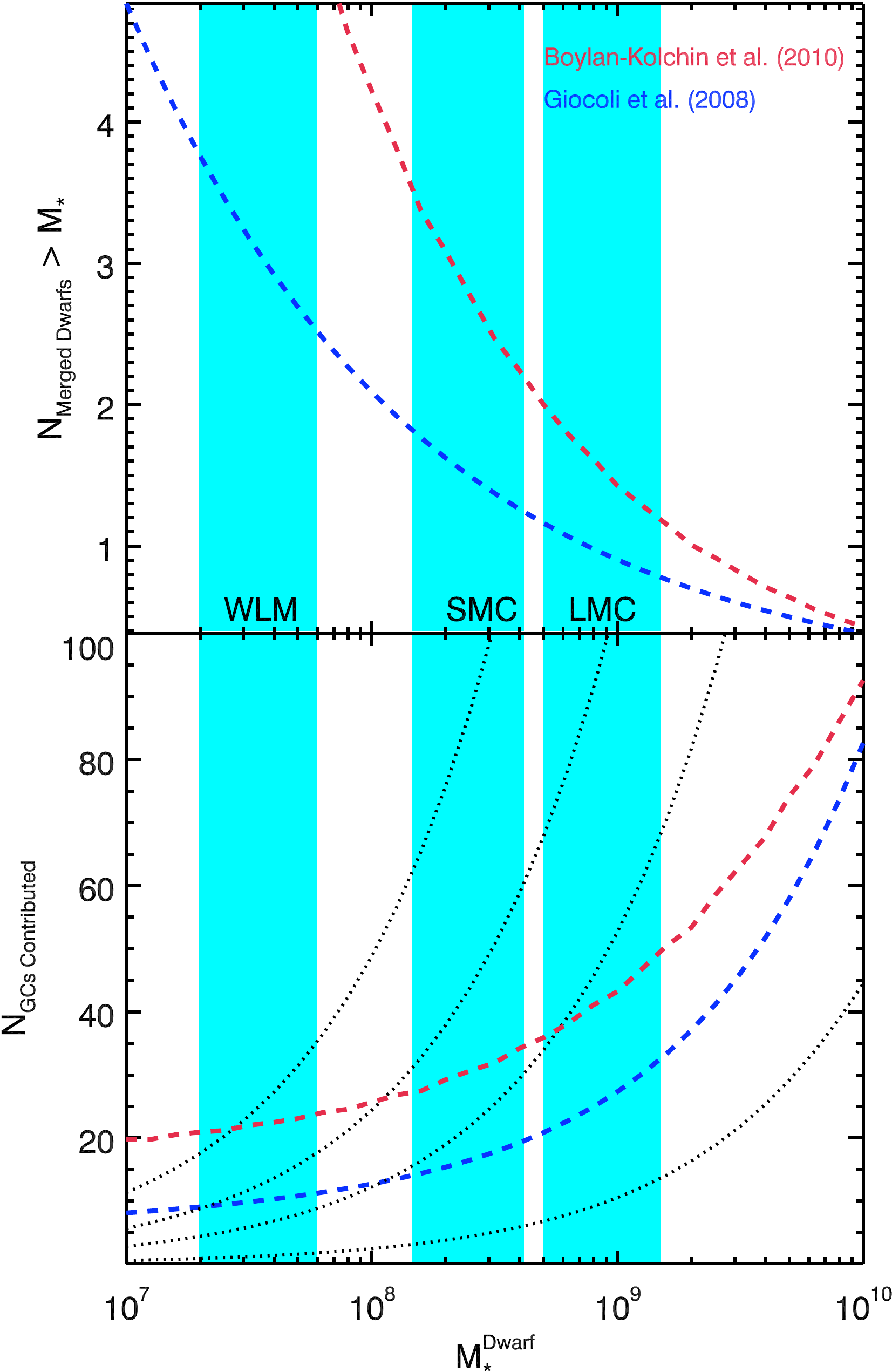}
\else
\includegraphics[width=0.49\textwidth]{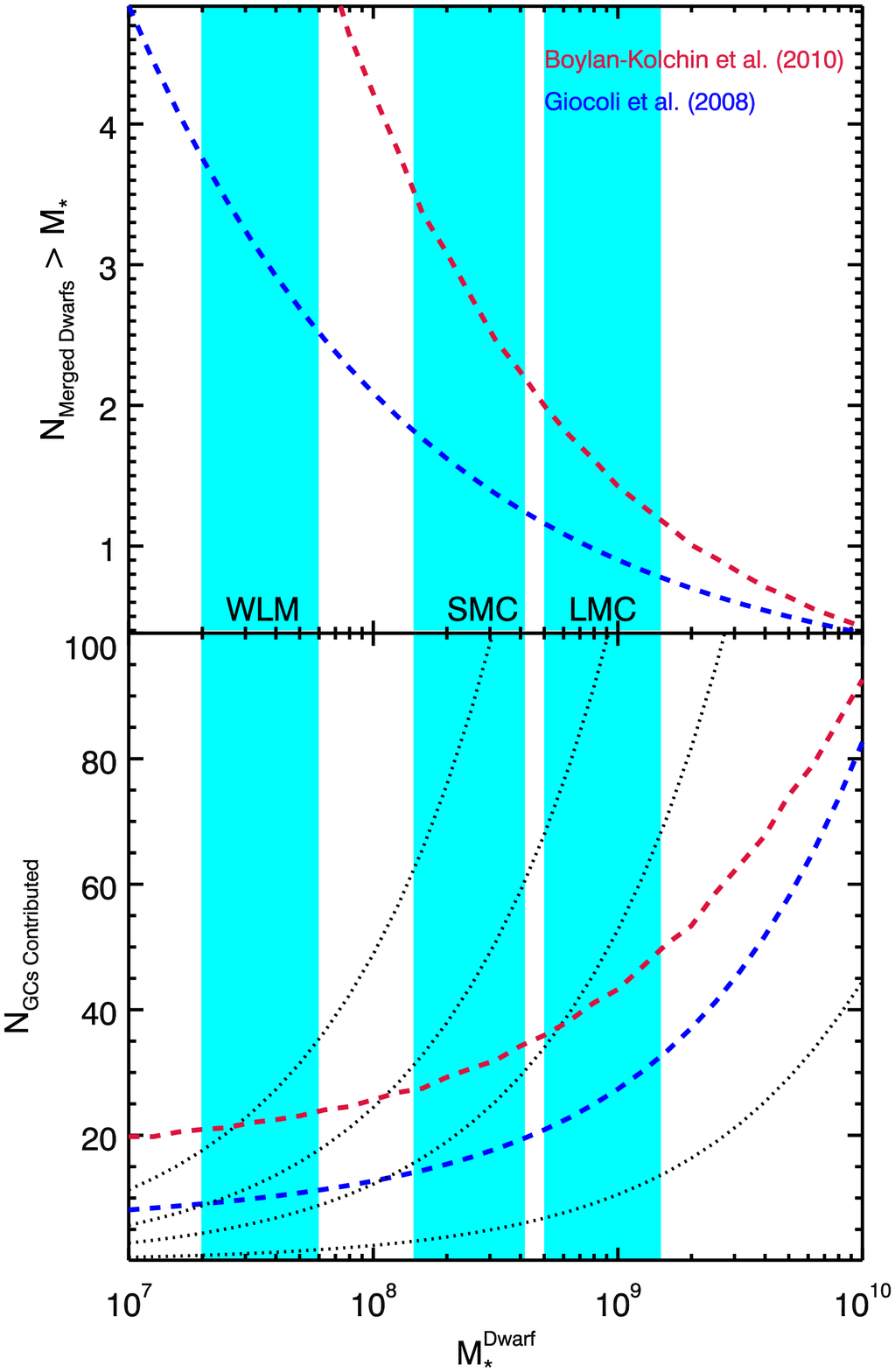}
\fi
\caption{Top panel plots the number of accreted dwarfs of a given stellar
mass expected to contribute to the MW halo over its buildup.  Bottom panel
plots, for various specific frequencies of $S_{N} = 20,10,5,1$
(\emph{dotted black lines}), what would be the total number of accreted GCs
from dwarfs of that mass. The blue and red dashed lines show the mass-dependent
$S_{N}$ relation from the previous figure.  Blue bands show for reference
where the Local Group dwarf galaxies would fall on these relations 
given their observed stellar masses \citep{Alan12}.}
\label{fig:sngc}
\end{center}
\end{figure}

\section{Birth Environments of the Accreted Halo GCs}
To help identify which galaxies could be responsible for the majority of the
accreted GCs (instead of just the cumulative number), Figure \ref{fig:sngc}
plots the number of merged dwarfs as a function of the stellar mass of the
dwarf, along with the total number of GCs which would be contributed by dwarfs
of that mass.  As in the previous figure, the number of GCs contributed by
objects of a given stellar mass is defined as $N_{GCs,contributed}(m_{*}) =
N_{GCs}(m_{*}) \times N_{merged}(m_{*})$, where $m_{*}$ is the stellar mass of
the dwarf.  The same curves of various specific frequencies
(20, 10, 5, 1, and the \cite{Peng08} $S_{N}$ relation for the two subhalo
distributions) are overlaid and we indicate where the three dwarf
galaxies that were considered in the AMR plots in the previous section would
fall on these curves.  

These predictions are based on the assumed merger history from the simulations
we consider, but suggest that if the MW has experienced 3 WLM-sized mergers, they
would have contributed only a total of $\sim 4$ GCs, while one merger of an
LMC-sized system would have contributed $\sim 35$ GCs\footnote{While the 
LMC itself does not contain this many \emph{old} GCs, dwarfs of similar mass in 
\cite{Peng08} show a wide range of $S_{N}$ values.}.  This is in close
agreement with positions of the halo GCs with respect to the AMRs of 
the LMC and WLM in Figure \ref{fig:gcamrs}.  Therefore it is
plausible that most of the Halo GC system came from $\sim 6-7$ mergers of
WLM- to LMC-sized dwarfs.  It is interesting to note that, while the absolute
number of merged dwarfs is different for the \cite{Giocoli08} or \cite{BK10}
subhalo mass distributions, they both suggest that the GC systems were
primarily built from the 6-7 most massive mergers.  We also note the good
agreement with the early estimates from \cite{Unavane96}, and again the 
simulations of \cite{Cooper10} who found that MW sized galaxies could accrete
$20-80$\% of their stellar halo mass from the most massive progenitor dwarf galaxy 
in the mass range of $10^{7} - 10^{8.5}$.

\begin{figure}
\begin{center}
\ifpdf
\includegraphics[width=0.49\textwidth]{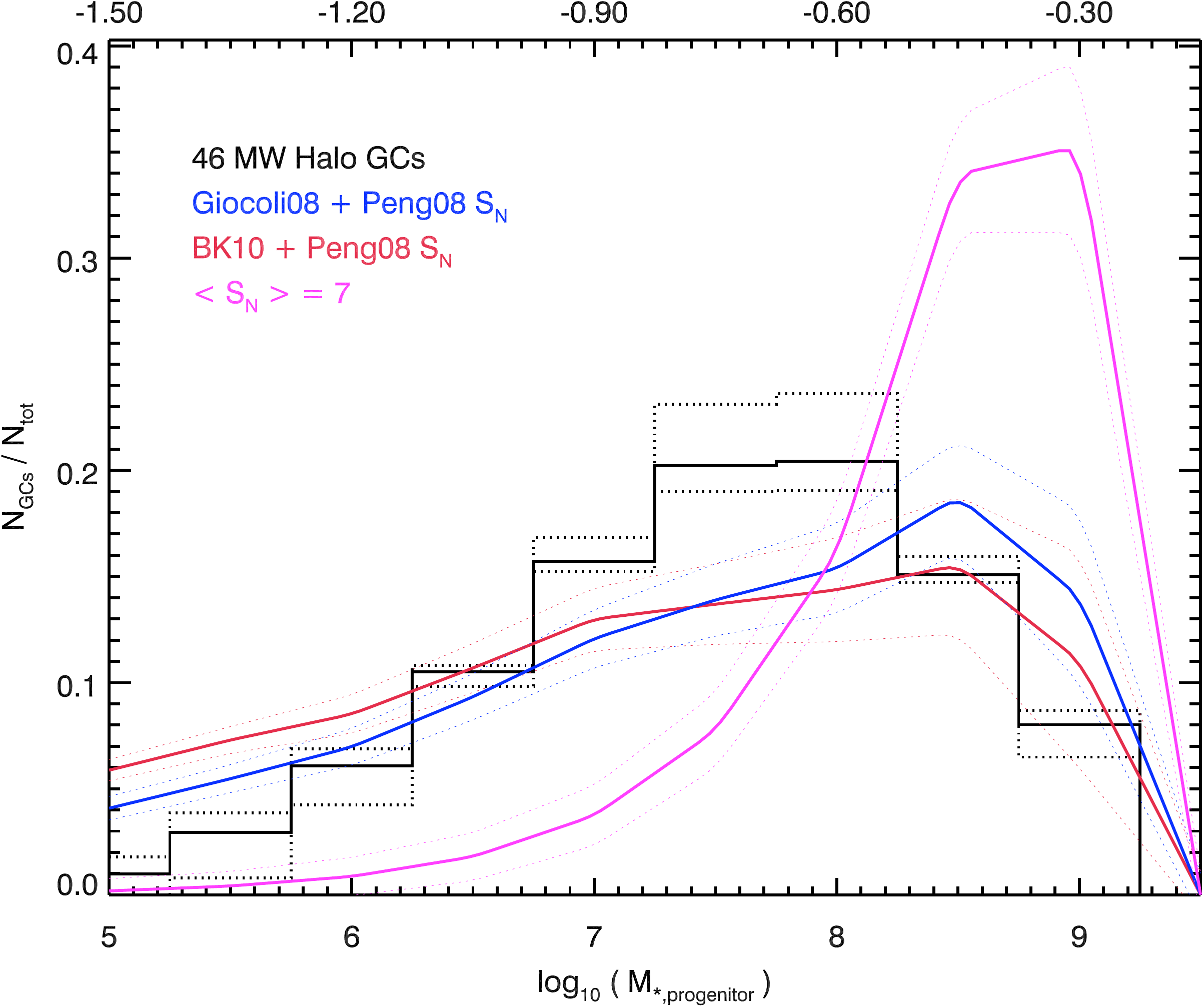}
\else
\includegraphics[width=0.40\textwidth,angle=-90]{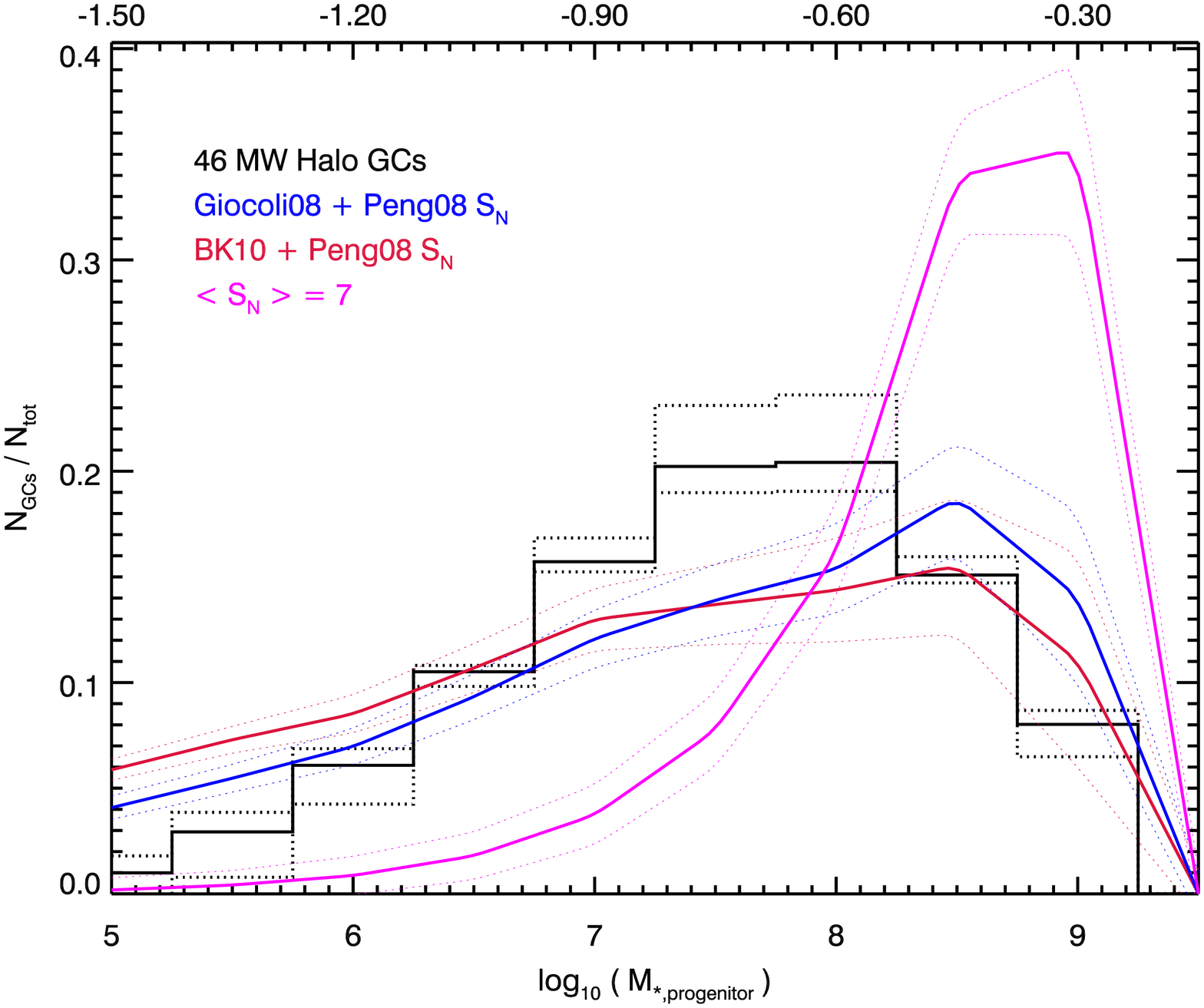}
\fi
\caption{Progenitor mass of the dwarf galaxies from which the observed MW halo GCs
could have formed, assuming the host galaxies offset AMRs translate
directly into a mass decrement with respect to the MW disk.  Top axis
shows the numerical value of the AMR offset based on the slope of the 
MMR. Black histogram
shows the result of applying this analysis to the 43 GCs classified as
belonging to the MW halo.  Red and blue dashed lines show the expected GC
contribution as a function of progenitor mass using the two subhalo mass
functions in the previous figures, and magenta line using a fixed $S_{N}$.}
\label{fig:mprogs}
\end{center}
\end{figure}

\subsection{Implied Masses of GC Progenitor Galaxies from the Mass-Metallicity Relation
and Offset AMR}
A more direct comparison of the progenitor galaxy masses of the MW halo
GCs with these model predictions, is possible by
again leveraging the mass-metallicity
relation in conjunction with the observed AMR.  We begin by modelling the 
metal-rich branch of the AMR (that associated with the in-situ disk clusters)
as a linear fit between the points ([Fe/H], Age)$ = $($-1.5, 13.0$) and ($-0.5,
10.75$).  From the near constant offset in metallicity at all ages for the disk
GCs (and similar shape of the observed dwarf galaxy AMRs), 
we can, to first order, assume that the slope of all of the AMRs in this
early epoch is invariant.  Any dwarf galaxy will then have an AMR that is simply
offset in metallicity from the disk AMR, by an amount which depends on the
total stellar mass of the dwarf.  

For each halo GC, (the black dots in Figure \ref{fig:amrprob}), we compute its
metallicity offset from the disk AMR, $\Delta\rm{[Fe/H]}$.  With the observed
slope of the mass-metallicity relation from \cite{Kirby11}, the value of
$\Delta\rm{[Fe/H]}$ for each GC is transformed into a mass decrement from the
MW disk, thereby giving a coarse estimate of the mass of the progenitor galaxy
in which that GC might have formed.  This is obviously most useful in a differential
sense, as the exact shape of the AMR and disk mass as the MW evolves is not precisely
known.

Figure \ref{fig:mprogs} displays the result of this analysis, wherein
we show the average distribution of implied progenitor dwarf galaxy masses for all 46 MW halo
GCs.  This is essentially a convolution of the subhalo mass function with the 
number of GCs per dwarf galaxy of a given mass. The black histogram represents the 
average distribution found from 50000 Monte Carlo trials of this process, in which
the observational errors in age and metallicity for the GCs, as well as the errors
on the fit to the disk GC AMR were simultaneously incorporated.

To directly compare these to the expected distribution from
simulations of the MW accretion history, we compute Monte Carlo realizations
as follows.  For a given trial we randomly pick 50 dwarf galaxies to merge 
with the MW with their distribution in masses weighted according to the 
subhalo mass function of Equation 3 (or Equation 7 of \cite{BK10}).
We take a lower limit for the progenitor galaxy mass as $M_{*} = 10^{5}$,
as that is a mass close to those of most GCs themselves, and the maximum progenitor
galaxy mass as $M_{*} = 10^{9}$, as we should not expect mergers greater than
the total MW stellar halo mass.  
The number of GCs per progenitor galaxy is computed for 
each of the 50 dwarfs assuming either the 
\cite{Peng08} $S_{N}$, or a fixed $<S_{N}>$ = 7, producing a predicted
distribution of the number of accreted GCs as a function of 
the mass of their progenitor galaxy.  This is repeated for 5000 trials,
and the mean and standard deviation of the distributions recorded --- which 
we show as the coloured curves in Figure \ref{fig:mprogs}.  

There is good agreement between the black histogram,
which shows the estimated progenitor masses
implied by the offset AMRs, and these
expectations from n-body simulations of the MW's assembly.  Dwarfs with
masses $\leq 10^{6} \msol$ contribute only a few percent of the total MW halo GCs,
with the contribution increasing until a maximum contribution is
reached, from  dwarfs of $\sim 10^{8.5} \msol$.  Notably, the 
predicted distribution when assuming a constant $S_{N}$ for all dwarf galaxies
results in far too many GCs from high mass progenitors.  This would suggest
that there needs to be at least some variation in the $S_{N}$ values of
the accreted system --- therefore the mass dependent variation of $S_{N}$ 
observed in the Virgo cluster sample of \cite{Peng08} may be representative
of many systems\footnote{We note also that the differential distribution of 
progenitor masses we recover using the \cite{Spitler09} relations is nearly identical
to what we find using the \cite{Peng08} formalism. However as the former
relation requires an assumption on the mass of individual GCs it provides
little leverage on the total number of GCs accreted (i.e. Figure 5 and 6).}
.  

The distribution of implied progenitor masses shown by the black histogram
allows us to study how many significant progenitor dwarfs built up the
total stellar halo and GC system following the definition in \cite{Cooper13}
\begin{equation}
N_{sig} = \frac{(\Sigma m_{*,progenitor})^{2}}{\Sigma m_{*,progenitor}^{2}}.
\end{equation}
We find a value of $N_{sig} = 11$, in good agreement with that 
work which showed that MW sized stellar halos
can be built up by $\sim 10$ significant progenitors in some cases.
Importantly, our semi-empirical estimate of $N_{sig}$ 
is completely independent of an
assumed subhalo mass function, as the progenitor masses are calculated solely
from the offset AMR, and provides a unique comparison to such simulations.
We note however that the exact 
distribution of the progenitor masses is somewhat sensitive to the 
relative shape of the MW and dwarf galaxy AMRs.  Assuming no pre-enrichment
in the MW disk AMR would lead to a suppression of the lowest mass progenitors
in the histogram, but would still not produce a peak at the very highest masses.

\subsection{Correlations between the Properties of GCs and their Implied 
Progenitor Galaxies}
\begin{figure*}
\begin{center}
\ifpdf
\includegraphics[width=0.81\textwidth]{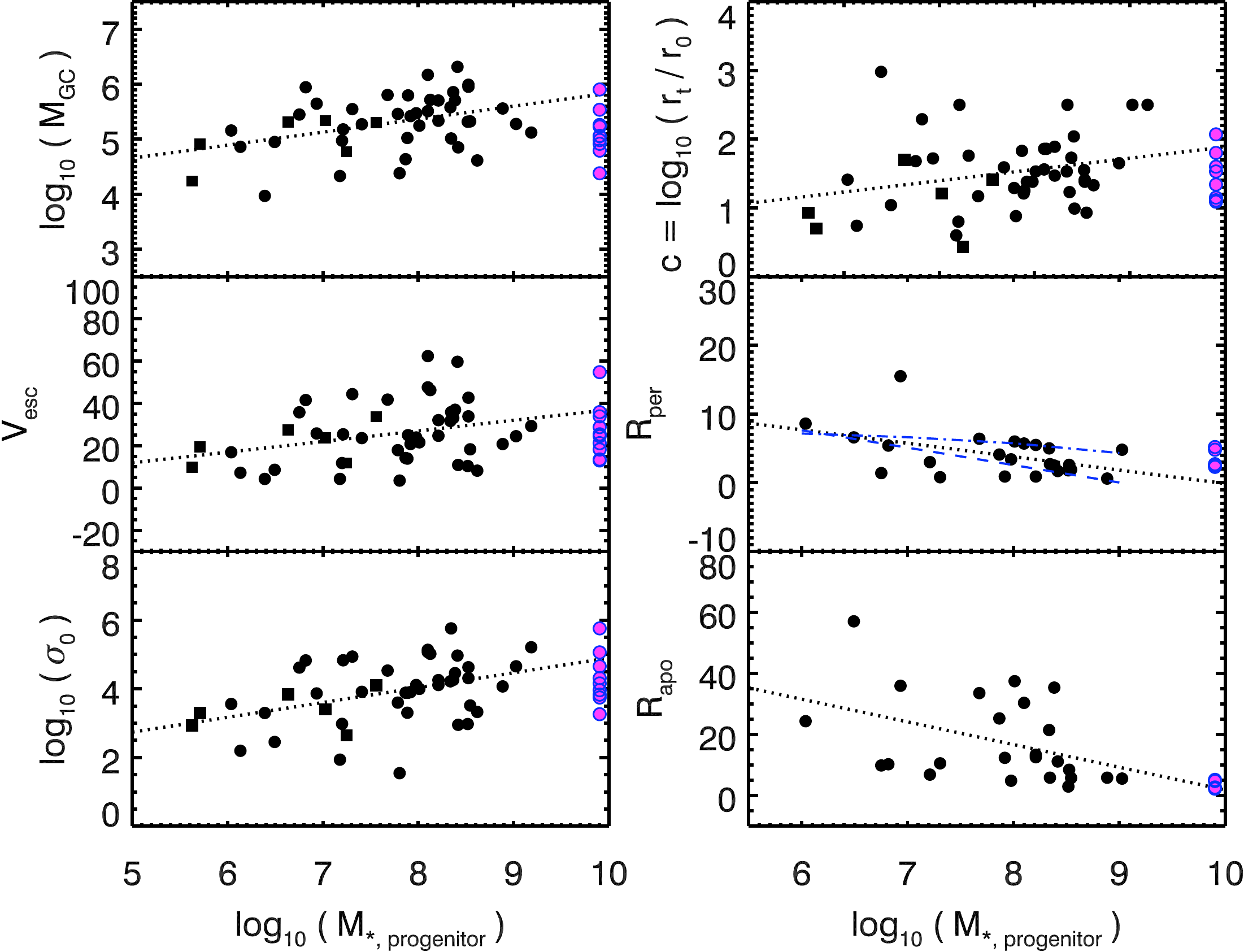}
\else
\includegraphics[width=0.62\textwidth,angle=-90]{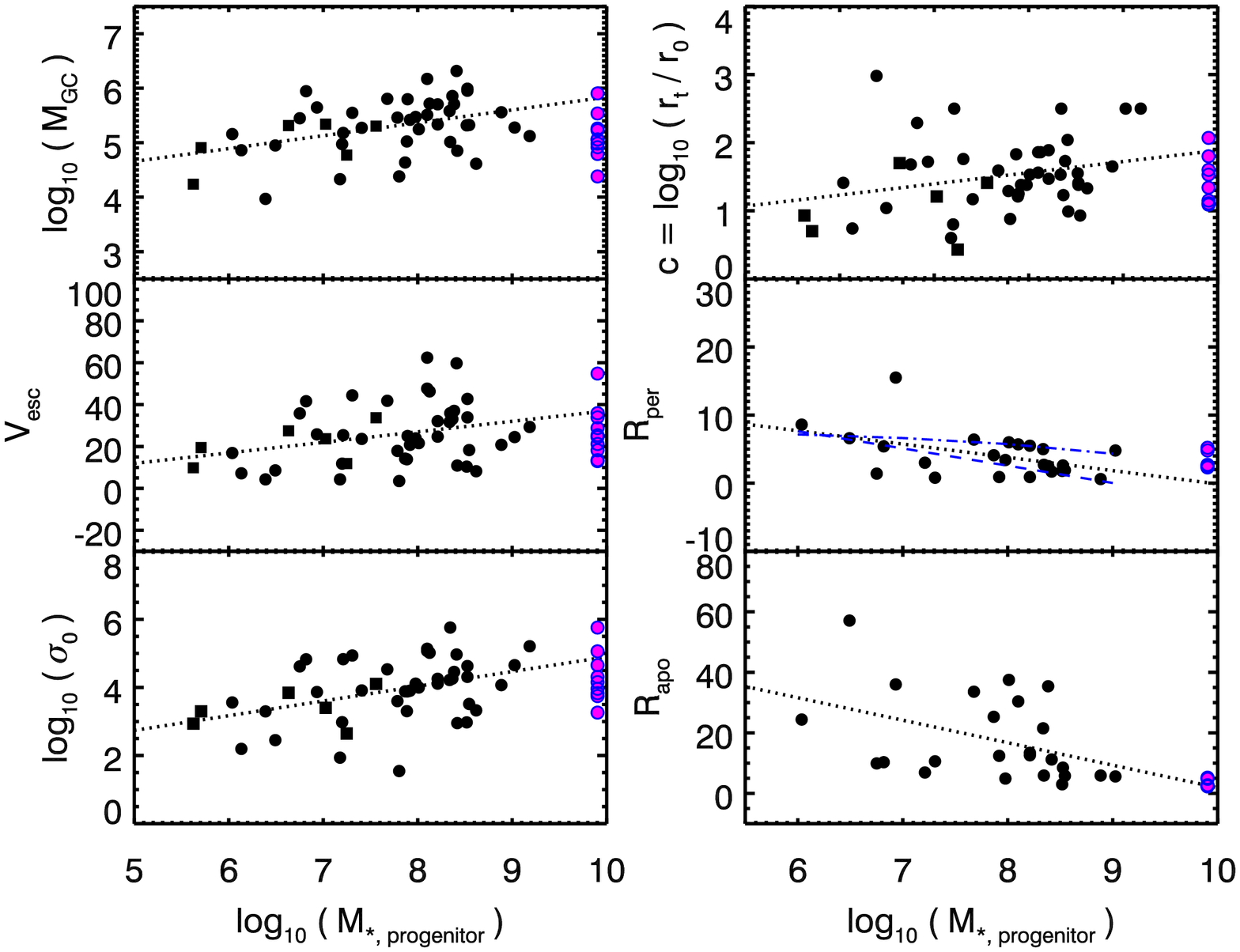}
\fi
\caption{Structural and orbital properties of the MW halo GCs as a function
of the mass of their progenitor dwarf galaxies.  Magenta points represent
the MW disk GCs.  Dotted lines are linear least squares fits to solely the
halo GCs.  Blue dashed lines show predicted orbital decay 
as a function of dwarf galaxy mass due to dynamical friction (see text).
}
\label{fig:progprops}
\end{center}
\end{figure*}

To further illuminate the link between the MW GC system and the host galaxies
of accreted halo GCs,
Figure \ref{fig:progprops} plots the implied progenitor galaxy masses
versus several structural and orbital properties of the globular clusters.
The disk GCs are shown as the magenta points and in this exercise are placed
at a progenitor mass close to that of the MW disk.  Linear least squares
fits are shown as dotted lines, and in all cases are computed based solely
on the halo clusters.

One noticeable trend is that the mass of a GC and its central density
(traced by $V_{esc}$ and $\sigma_{0}$) are correlated with the mass of the
galaxy in which the GC was born.  While speculative, these correlations may
provide an explanation as to why GCs with the highest $V_{esc}$ and $\sigma_{0}$
in V13 tended to show the strongest evidence for helium abundance enhancements
(as inferred from the relatively steep slopes of their subgiant branches in the
observed CMDs).  This result is particularly puzzling, as the \emph{present-day}
cluster masses should not be predictive of past masses and their capability, or
not, to retain the ejecta from a primordial stellar generation and to produce
second-generation stars with significant He abundance variations.  Figure
\ref{fig:progprops} suggests that the most massive, densest clusters may form
predominantly in the highest mass dwarf galaxies (although a range in $V_{esc}$
and $\sigma_{0}$ is apparent at any progenitor mass, suggesting perhaps that
even within a single dwarf, GCs and molecular cloud mass may have a dependence
on local environment; e.g., \citealt{Meidt13}).

Therefore, when they are accreted by the MW, those GCs would be embedded in a
much larger dark matter and baryonic envelope which could mitigate the effects
of tides and ram pressure, perhaps allowing them to retain ejecta from their first
generation of stars.  

The correlation between concentration parameter 
($c = log_{10}(r_{t}/r_{0})$; \citealt{Harris10}) and
progenitor mass in the top right panel of Figure \ref{fig:progprops} is 
particularly interesting in light of the correlation between GC size
and colour (or metallicity) seen in the MW and many 
extragalactic GC systems in giant ellipticals (e.g., \citealt{Larsen01}).  As the 
correlation persists over many radii within individual galaxies, \cite{Larsen01}
suggested that this may reflect a primordial correlation which is setup 
in the birth of the GC.  They suggested that perhaps the most compact GCs
form in the largest gas clouds (which would be found in the largest galaxies)
preferentially.  This interpretation would seem to be supported by the trend 
in the top right panel of Figure \ref{fig:progprops}, and would suggest that
the correlation exhibited between concentration and colour or metallicity
is a consequence of the lower mass progenitor systems in which preferentially
low metallicity systems (of a given age) are found.

The bottom two right-hand panels of Figure \ref{fig:progprops}
show that there is also a
trend for GCs with the lowest host galaxy masses to be found at larger
Galactocentric distances.  This correlation may arise as a consequence of the
orbital decay of a dwarf galaxy as it is accreted onto the MW.  
Galaxies with the highest masses (and the GCs they host) will preferentially
sink to the centre of the MW via dynamical friction.  This effect is enhanced
further when the mass growth of the MW is taken into account or for more
eccentric orbits\footnote{However as shown by \cite{vandenbosch99}, there is no
total evolution in orbital eccentricity, which likely explains why the slopes
of the pericentric and apocentric radii relations are so similar.}.  We note,
however, that there is no strong evidence from simulations for systematic
variations in the radius as a function of the progenitor mass \citep{Wang11}.

An order of magnitude estimate for the varying impact of dynamical friction
on dwarfs of different masses can be estimated from the analytic expressions
given in \cite{Zhao04}.  These formulae for dynamical friction consider both
tidal stripping of the accreted satellite, and growth of the MW halo over time,
and the change in orbital radius with time is characterized by 2-parameter 
functions of the form:
\begin{equation}
\Delta r(t) \simeq \frac{2{\pi}Gt_{f}}{(t_{f}-t_{i})V_{f}}m_{i}[1 - (1 - (\frac{m_{f}}{m_{i}})^{1/n})]^{n}\int^{t_{f}}_{t_{i}}t^{n-p}dt
\end{equation}
We compute this quantity for dwarfs of initial mass $m_{i} = 10^{6},10^{7},10^{8},10^{9}$,
and track the dwarfs orbital decay from $t_{i} = 4$ Gyr to $t_{f} = 13$ Gyr where 
the remnant accreted dwarf mass is reduced to $m_{f} = 10^{5}$.  We then compare
the relative orbital decay for each of the different mass dwarfs, allowing for comparison
with the right hand panels of Figure \ref{fig:progprops}.

A reasonable reproduction of the observational data is found for
a mass loss parameter value of $n = 1.01$ (corresponding to a near linear case, 
suitable for isothermal dwarf galaxy models) and
qualitatively reproduces the trend of radius versus dwarf galaxy progenitor
mass seen in the right hand panels of Figure \ref{fig:progprops}.
  We also show the formulation given in 
\cite{Cote98} which has a milder slope due to the fact that it doesn't 
incorporate mass loss for the host systems or growth of the MW. 

\begin{figure}
\begin{center}
\ifpdf
\includegraphics[width=0.49\textwidth]{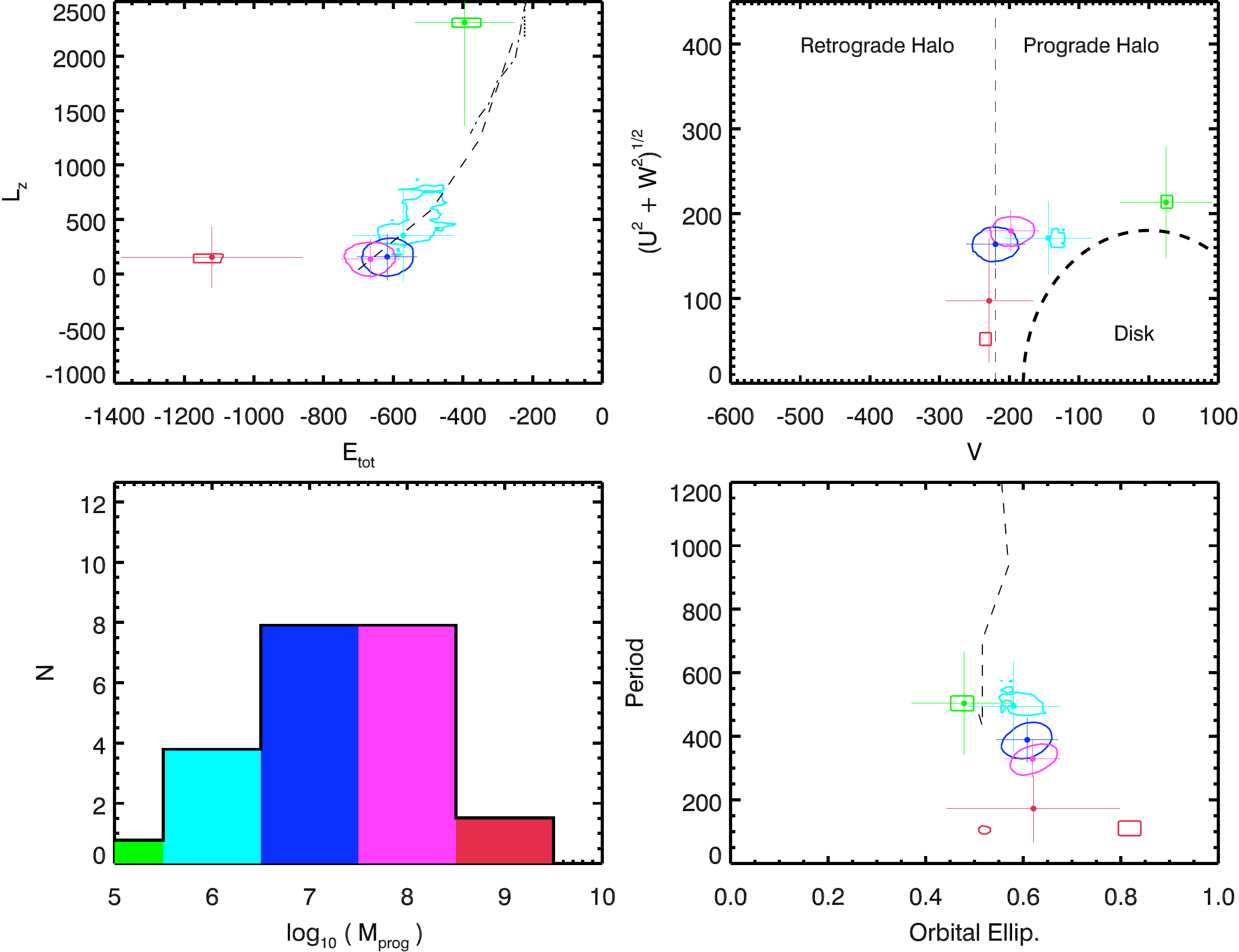}
\else
\includegraphics[width=0.40\textwidth,angle=-90]{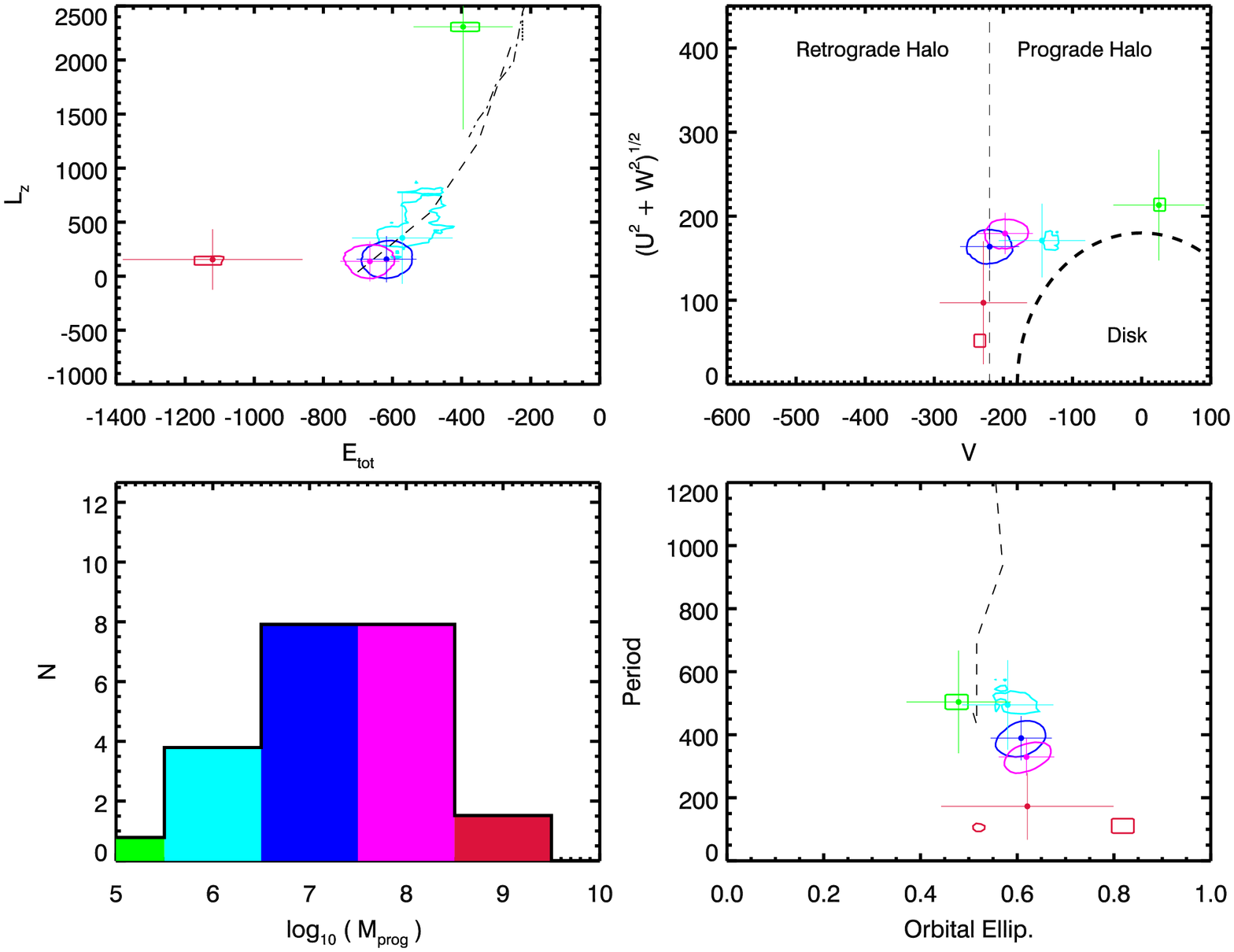}
\fi
\caption{Monte Carlo simulations of orbital properties for MW halo GCs.
Bottom left panel shows histogram of progenitor masses after 50000 trials
where the cluster progenitor masses are computed taking into account errors
on their age, metallicity and AMR of the MW disk.  The histogram bins are
colour coded by mass, and the other three panels show
the $1\,\sigma$\ contours, and median distribution values (\emph{crosses}) of
each mass bin for various orbital parameter spaces.  There is a clear
tendency for the GCs to separate by progenitor mass in their orbital
properties.  Lines in the top left and bottom 
right panels show expectations for the
impact of dynamical friction from \protect{\citep{vandenbosch99}}.}
\label{fig:kmeansamr}
\end{center}
\end{figure}

A further correlation between the orbital properties of GCs of 
similar progenitor mass is shown in Figure \ref{fig:kmeansamr}.  The cluster
progenitor masses are computed for each cluster for 50000 Monte Carlo trials,
with the errors on age, metallicity and disk AMR incorporated as described previously.
For each trial the clusters are grouped by progenitor mass and the 
orbital properties of the clusters in each mass bin are recorded.
  The panels in Figure \ref{fig:kmeansamr} show that \emph{over many
realizations} the GCs from different progenitor masses tend to show 
common and distinct orbital properties.  The GCs from the highest mass clusters
again show properties that are consistent with having been operated on by dynamical
friction.  The mild separation of clusters by their progenitor mass 
in Figure \ref{fig:kmeansamr} provides some support that our simple 
estimates for progenitor mass are physically consistent with their phase
space properties in this context.  
We stress however that there are relatively large, yet unquantified systematic 
errors associated with the derived orbital quantities used in this Figure.
Similarly, the particular separations by progenitor mass in this Figure are not
expected to be a generic prediction for all galaxies, as the variety
in merger history is large; however more detailed orbital models of individual
GCs in the MW or external galaxies could continue to explore such 
clustering of GCs in phase space in this way.

\section{Discussion}
\subsection{What Controls the Range in GC Ages in the MW}
It is well known that the MW GC population shows a sharp truncation in their
relative ages, with no GCs younger than $\sim 8$ Gyr \citep{vdb00}.  The
in-situ disk clusters, and even the more metal-rich of the halo clusters (which
would have come from the most massive accreted dwarf galaxies) have GCs that
span the first $2-4$ Gyr of the MW's history.  The more metal poor GCs
(corresponding to the lower mass WLM-sized dwarfs) do not contain similarly
young clusters (note the deficit of points around $-2.0$, and 11.25 Gyr in Figure
\ref{fig:amrprob})\footnote{The notable exception is Ruprecht 106 at 
[Fe/H]$=-1.78$ and 10.75 Gyr in this plot, however as discussed in $\S 2$ 
the metallicity of this cluster is particularly uncertain, with \cite{Dotter11}
favouring a value of [Fe/H]$=-1.5$.}.

Indeed WLM itself, with a stellar mass of $\sim 5 \times 10^{7} \msol$, has only one
old GC, while the LMC and SMC, and particularly Sagittarius are known to have
extended cluster AMRs of 3 Gyrs or more \citep{Glatt08,Law10,Colucci11}.  This could be an
indication that less massive dwarfs, which exhaust their gas reservoirs quickly
(perhaps due to the efficient SNe induced blowout of gas from their shallow
potential wells), form GCs for only a short interval of time.  Alternatively,
it may be the case that their SF efficiencies are so low,
or that they are more susceptible to reionization (e.g. \citealt{Strader04})
so that few stars (and GCs), are able to form before they are accreted.  

A complementary picture could be that the lowest mass dwarfs, in which 
relatively few GCs formed due to reasons given above, were accreted earlier by
the MW.  This would effectively truncate their cluster AMR sooner than in the
case of higher mass dwarfs which were accreted at later times by the MW, and
therefore had time to form GCs over a longer epoch.  This is supported by
simulations of MW-sized halos, where the average accretion time of low-mass
dwarfs is found to occur earlier by $\sim 4$ Gyr \citep{BK10}.  
If massive GCs are necessarily formed 
in very intense starbursts \citep{Elmegreen12} or massive
GMC collisions \citep{Furukawa09}, it could simply mean that these conditions are not found
in dwarf galaxies or the MW (which shows a relatively quiescent SFH).

Such a scenario might also explain why the MW is devoid of GCs younger than
$\sim 8$ Gyrs, with even the most metal-rich halo GCs having older ages.  In
order to have an increasingly longer GC formation epoch, a dwarf would require
both a late time accretion to the MW and be a comparable fraction of the MW
mass.  As such high-mass mergers are increasingly unlikely in CDM cosmologies,
the lack of GCs younger than $\sim 8$ Gyrs may simply be telling us that there
has not been any accretion of dwarfs with stellar masses greater than
$10^{9} \msol$.

\subsection{Formation Theories for the 
MW GC System in the Context of this AMR}
The primary theories for the assembly of GC systems 
in large galaxies (see review by \citealt{Brodie06}) are:
accretion of GCs in dwarf galaxies 
during hierarchical assembly \citep{Cote98},
merging of large galaxies \citep{Ashman92},
and in-situ formation \citep{Forbes97}.
Observational data for MW and extragalactic systems has revealed that
most GC systems are comprised of a red and a blue population which
are generally considered to reflect the underlying metallicities
of the GCs (although see \citealt{Yoon11,Blakeslee12}).
Correlations between the mean colour 
of the red, and blue GC populations with host galaxy mass have been used
to help understand possible formation avenues.

This is intrinsically challenging given the relatively
unknown age distribution
of the blue GCs in extragalactic systems, which makes inferences
concerning the formation time, duration and mechanism of in-situ or accreted
GCs difficult.  The analysis presented here suggests that an accretion
formation for the metal poor (halo) GCs in the MW is supported by
the details of the AMR shape with respect to the disk clusters.
Much of our analysis is in line with the earlier proposals from
\cite{Cote98}, and the primary tension of that work with observations 
--- that the predicted red and blue populations of GCs show no mean
age difference --- is no longer inconsistent with this MW GC AMR.
Similarly the work of \cite{Muratov10} which incorporated many
aspects in agreement with our analysis, is not problematic in 
its suggestion that large age spreads are likely in both the
metal poor (MP) and metal rich (MR) GC populations 
(with that study also finding a similar ratio of in-situ disk clusters 
as seen here).
Additionally the observation that GC metallicities in dwarf galaxies
are lower at all ages than in higher mass galaxies \citep{Strader05}
is not a point of concern with accretion models, 
but is a natural expectation given the
mass metallicity relation for the accreted (dwarf) galaxies.

Alternatively in-situ theories, which were envisioned to explain
blue GC populations with a narrow range of old ages \citep{Forbes97,
Beasley02}, typically find that the MP GCs should be older by $\geq 2$
 Gyr than the MR GCs and have a very narrow range of ages.
In some cases the in-situ theory (e.g., \cite{Forbes97}) 
produces a narrow range of ages for the MR component as well,
as the dynamical time for the galaxy during its secondary collapse
has decreased.
To first order both aspects of these theories are ruled out for the 
MW, as the MR and MP AMR sequences have identical average ages, and
age spreads.
 
Hybrid theories \citep{Strader05} may have less tension with the 
MR AMR but suggest that the blue GC population is not assembled
via hierarchical accretion. This new AMR does not invalidate such 
theories in principle, however the suggested high redshift, rapid formation
epoch for blue GCs is again problematic in light of the halo GC
age range of the MW.  The main argument in \cite{Strader04} 
against accretion of DGs to assemble the blue GC population is that 
it would tend to erase the correlation between average blue GC colour
and host galaxy luminosity.  This may not necessarily be
the case when one considers that the GCs are pre-enriched
from dwarfs that themselves fall on the mass-metallicity relation
for galaxies.  Therefore the GC population of the 
dwarf galaxies reflects the self-enriched history of
their birth environment \citep{Leaman12}, and a \emph{hierarchical}
accretion scenario (see e.g., Fig. 14 of \citealt{Cooper13})
would also produce a correlation 
between the metallicity of the blue GC population
and host galaxy metallicity.

Certainly in-situ or hybrid formation models which don't truncate 
the epoch of formation for blue GCs may be possible, however
it could be difficult to produce two in-situ populations of GCs
with the same age distributions but offset in metallicity within 
a single galaxy.

\subsection{Prediction for the AMR of the MW Bulge Globular Clusters}
V13 did not include any of the highly reddened clusters thought to be associated
with the bulge of the MW in their sample.  These clusters, in particular,
provide a test of how the central bulge of the MW formed; i.e., via mergers or
through secular evolution, and therefore their ages and metallicities can be
used to probe these two scenarios.  Our interpretation of Figure
\ref{fig:gcamrs} indicates that the metallicity offset between the disk and
halo branches of the AMR is driven by the relative mass of the self-enriching
environment (Milky Way disk, or SMC sized dwarf galaxy) in which the clusters
formed.  

In this context, the MW bulge as a distinct environment with a unique chemical
enrichment history (as compared with the disk), should have clusters that 
formed in-situ which follow an AMR that is proportional to the stellar mass of
the bulge.  Mass models from \cite{McMillan11} give the bulge stellar mass as
$\sim 9\times 10^{9} \msol$ with earlier studies quoted in that paper favouring 
higher masses of $(2.4 \pm 0.6) \times 10^{10} \msol$.
Given the stellar disk mass of $\sim  (3 \pm 1) \times 10^{10} \msol$, 
the small offset in stellar mass between the two
components would suggest that the bulge clusters of a given age have 
close to the same metallicities as disk clusters of the same age, or are nearly
coeval with disk clusters at a given [Fe/H] value.  This region is shown as the
dashed green polygon in Figure \ref{fig:gcamrs}

Using recently measured spectroscopic metallicities for bulge clusters from
\cite{Saviane12} and others in the Harris (2006) GC
catalogue\footnote{http://physwww.mcmaster.ca/~harris/mwgc.dat} we might
predict the ages of some of the bulge GCs (under the assumption they formed
in-situ).  The bulge clusters near $\rm{[Fe/H]} \simeq -1.35$
(including HP 1, AL 3, NGC 6522, NGC 6540, Terzan 4, and NGC 6325 as well as
more metal poor clusters) are expected to have ages of $\sim 12.75$ Gyr.
Similarly, it is our expectation that clusters with [Fe/H] $\sim -1.05$
(e.g., NGC 6558, Terzan 9), $\sim -0.5$ (NGC 6356, NGC 6441), and 
$\sim -0.2$ (NGC 6528, NGC 6553, NGC 6440) will have ages of approximately 
12.0, 10.75, and 10.0 Gyr, respectively.  Due to the
difficulties in determining absolute cluster ages, the prediction most easily
tested is that the ages of NGC 6558 and Terzan 9 are expected to be $\sim 1.5$--2
Gyr older than NGC 6528/6553/6440.

As shown by \cite{Cote98}, such analyses can also be applied to more distant
galaxies (M31, or the GC systems of elliptical galaxies) in order to place
constraints on the relative number of accreted GCs, and to achieve a better
understanding of the formation environments of GCs.  

\subsection{Caveats}
In this analysis we have assumed that the entire halo population of GCs and
MW stars were formed through accretion.  There is conflicting observational
evidence in the literature as to whether the halo of the MW shows two distinct
components \citep{Beers12,Schonrich11}, and if so, whether one of the components
is incompatible with an accretion origin.  Certainly there is
theoretical work suggesting that some fraction of the MW halo stars may have formed
in-situ from high-redshift accreted gas, and then been dynamically heated into the
MW halo through later mergers (e.g., \citealt{Zolotov09}).  Such simulations
have found that the in-situ fraction strongly depends on the merger history of
the MW.  Unfortunately it is not known whether any GCs would also form in-situ
through the same process; therefore, we simply note that the assumed accreted
stellar mass of the MW halo can be considered an upper limit.  Any contribution
by stars formed in-situ would then lower the number of merged dwarfs (and hence GCs)
in Figure \ref{fig:sngc2}.  Whether or not the observed GC fraction could be
used to constrain the in-situ halo fraction in this way is likely not possible
with this analysis however. 

Similarly, higher-order chemical abundances ($\alpha-$, $r-$ and $s-$process
element ratios) have the ability to discern whether the chemical evolution of
the GCs and halo field stars are similar.  Results are complicated by the
unknown formation epoch and environment for GCs.   However, there is evidence
in the MW, LMC, and Fornax that the metal-poor GCs show similar [$\alpha$/Fe]
ratios as in the field stars that have similar [Fe/H] values
(\citep{Pritzl05,Hill97,Letarte06,Mucciarelli11}, but see \citealt{Mateluna12}).
This helps to alleviate one concern with the present analysis, in that we do
not consider what fraction of GCs may have evaporated during the accretion
procedure.  While there may have been many small GCs which were disrupted and
dispersed through the halo during accretion, we can consider them part of the
stellar mass contributed by dwarfs in light of the above results.

The same chemical similarity is not true of the MW halo stars and the field
stars in the surviving dSph galaxies \citep{Venn04}.  However we note that this
is not particularly constraining for the current problem as the dwarf galaxies
studied are the \emph{surviving} population of dSphs, which may have very
different chemical properties from the ones that merged with the MW (c.f.,
\citealt{Gilmore98}), especially if the mergers with the MW happened early,
before significant chemical evolution of the dwarfs could have occurred.
Certainly, if the lowest mass dwarfs were accreted first, the [$\alpha$/Fe]
ratios could well be markedly different from what is observed in dwarfs today.
Additionally, Figure \ref{fig:gcamrs} suggests that high-mass dwarfs
($\sim 10^{8-9} \msol$) were the predominant contributers to the GC population
of the MW halo (see also \citealt{Cooper13}), in which case,
the chemical abundance differences between the
MW field stars and low-mass dSphs ($\sim 10^{6-7} \msol$) is not necessarily a
relevant constraint on this scenario.

Finally, a split AMR could generally be interpreted not only as an offset in 
metallicity at fixed age, but as an offset in age at fixed metallicity.  The 
latter might be a signature of a two-phase galaxy 
collapse (e.g., \citealt{Hartwick09}).
However given that this particular AMR split does not extend to the
lowest metallicities and the range in ages are identical, the data analyzed here
 seem to favour the scenario presented in $\S4$ over the interpretation of an age offset.

\section{Conclusions}
Our analysis of the AMR in the MW GC system has identified a clear split in the
sequence, with approximately one-third of the clusters being offset by $0.6$
dex in metallicity from the more populous metal-poor branch\footnote{The data for the AMR figures 
in this paper can be found in Table 1 of V13, 
and is also available upon request from the authors.}.  The corresponding
mass decrement as implied by the MMR, coupled with subhalo merging statistics
from simulations of MW-sized halos, has led us to postulate that the MW halo GC
system could have been assembled in a consistent manner simultaneously with the
MW stellar halo.  This interpretation was also suggested by \cite{Elmegreen12},
who inferred from the space density of Ly$\alpha$ emitting galaxies that some
metal-poor (halo) GCs formed in dwarfs with stellar masses comparable to WLM,
in agreement with our independent analysis.

A notable implication of the new AMR is that the
identical mean age, and spread in ages
shown by the metal rich disk GCs and the metal poor halo GCs is difficult
to reconcile with in-situ formation for the latter population.  The $\sim 4$ Gyr 
age spread amoung the MW's metal poor halo GCs could imply that the 
blue GCs in extragalactic galaxies need not have formed in a truncated epoch
due to reionization.  Similarly correlations between the average metallicity
of the blue GC population and the host galaxy mass is consistent with
the hierarchical accretion origin of the halo GCs.

The observational result of the split AMR is only made possible by the robust
age derivations from V13, which have revealed and corrected several systematics
in past analysis of the MW GC ages.  Past studies of the AMR of the MW GCs have
therefore been limited by the data in reaching conclusions concerning the
origin of various GC populations (e.g., \citealt{Forbes10}).

Our interpretation of the bifurcated AMR as a diagnostic for understanding the
birth environment for GCs, is also useful as a predictive tool for the MW bulge
population, and may offer valuable constraints on simulations that investigate
the amount of accreted stellar mass in the MW halo.  Further high fidelity age
and orbital properties for GC systems in the MW, M$\,$31, and other galaxies
will continue to aid such ventures.

\section*{Acknowledgments}
The authors thank the anonymous referee for useful comments which greatly
improved this manuscript.  
We thank Ata Sarajedini for informing us where to obtain the publicly available
photometry for five of the outer-halo GCs for which we have determined ages.
The authors also thank Kim Venn, Else Starkenburg, Alan McConnachie and
David Hartwick for several helpful comments, which improved this paper.  
RL acknowledges financial support to the DAGAL network from the
People Programme (Marie Curie Actions) of the European Union’s Seventh
Framework Programme FP7/2007- 2013/ under REA grant agreement number
PITN-GA-2011-289313, whereas the contribution of DAV was supported by a 
Discovery Grant from the Natural Sciences and Research Council of Canada.

\bibliography{gcamr.bib}


\label{lastpage}

\end{document}